\documentclass[twocolumn,showpacs,prl,amsmath,amssymb,floatfix]{revtex4-1}
\usepackage{amsmath,amssymb,color}
\usepackage{bm}
\usepackage{epsfig}
\usepackage{psfrag}
\usepackage{hyperref}

\newcommand{\bd}{\bm}

\begin{document}

\title{
Rayleigh-Jeans condensation of pumped magnons in thin film ferromagnets}

\author{Andreas R\"{u}ckriegel}
\affiliation{Institut f\"{u}r Theoretische Physik, Universit\"{a}t Frankfurt,  Max-von-Laue Strasse 1, 60438 Frankfurt, Germany}

\author{Peter Kopietz}
\affiliation{Institut f\"{u}r Theoretische Physik, Universit\"{a}t Frankfurt,  Max-von-Laue Strasse 1, 60438 Frankfurt, Germany}

\date{June 11, 2015}

 \begin{abstract}
We show that the formation 
of a magnon condensate in thin ferromagnetic films
can be explained within the framework of a classical stochastic non-Markovian
Landau-Lifshitz-Gilbert equation where the properties of the
random magnetic field and the dissipation are determined
by the underlying phonon dynamics.
We have numerically solved this equation
for a tangentially magnetized yttrium-iron garnet film in the 
presence of a parallel parametric pumping field. 
We obtain a complete description of all stages of the nonequilibrium time evolution
of the magnon gas
which is in excellent agreement with experiments.
Our calculation proves that the experimentally observed condensation
of magnons  in yttrium-iron garnet at room temperature 
is a purely classical phenomenon which should be called Rayleigh-Jeans 
rather than Bose-Einstein condensation.

\end{abstract}

\pacs{75.30.Ds, 75.10.Hk, 05.30.Jp}

\maketitle

In the past  decade the nonequilibrium dynamics of parametrically pumped 
magnons  in thin yttrium-iron garnet (YIG) films
has been investigated by many experimental 
studies \cite{Demokritov06,Demidov07,Dzyapko07,Demidov08a,Demidov08b,Demokritov08,Serga14,Clausen15a,Clausen15b}.
Very rich physics was found, including the 
overpopulation of the lowest energy state, 
which was interpreted as Bose-Einstein condensation (BEC) of  magnons 
at room temperature and finite momentum.
Using the technique of Brillouin light scattering it is even possible to measure the 
magnon distribution with momentum and time resolution \cite{Sandweg10}.
This allows an observation of the parametric resonance and of the subsequent thermalization leading to the formation of the condensate in detail \cite{Demidov08b,Serga14}.  
Unfortunately, a complete theoretical understanding 
of this phenomenon is still lacking and 
there is no theory that can simultaneously describe all stages of the experiment.
While the so-called 
S-theory \cite{Rezende09,Lvov94,Zakharov70,Tsukernik75,Vinikovetskii79,Lim88,Kalafati89,Kloss10} is able to describe the parametric resonance used to populate certain magnon states,
it does not properly take magnon-magnon scattering into account
 and therefore cannot describe the cascade of relaxation processes 
leading to the formation of a magnon condensate.
On the other hand, theories that focus on the condensate
 usually do not take the pumping dynamics into account and start with some given quasiequilibrium state which can be identified with the ground state of some effective
quantum mechanical Hamiltonian  \cite{Tupitsyn08,Hick10,Li13}. 
Phenomenological approaches of the Ginzburg-Landau type also have been used to study the 
condensation dynamics \cite{Malomed10}.
Finally, theories dealing with the relaxation processes and kinetics of excited 
magnons did not include the possibility of magnon 
condensation \cite{Lavrinenko81,Taranenko89,Lutovinov89,Hick12}.

Since BEC is a manifestation of quantum mechanics,
it seems at the first sight reasonable that 
quantized magnons obeying
Bose statistics are essential to obtain a  satisfactory theoretical
description of magnon condensation in YIG.
However, since the experiments are performed at 
room temperature, which is large compared with
the relevant magnon energies, the equilibrium distribution of the magnons
is the Rayleigh-Jeans rather than the Bose-Einstein distribution.
This suggests that the experimentally observed magnon condensation
at room temperature \cite{Demokritov06,Demidov07,Dzyapko07,Demidov08a,Demidov08b,Demokritov08,Serga14,Clausen15a,Clausen15b} is a purely 
classical phenomenon which should be called  Rayleigh-Jeans condensation
rather than BEC. 
The concept of Rayleigh-Jeans condensation has been discussed before by Kirton and Keeling in the context of condensation of photons \cite{Kirton15}.
In fact, it is well known that
classical waves subject to some randomness and mode-coupling
can undergo a kinetic condensation
transition analogous to BEC of quantum systems \cite{Connaughton05,Duering09,Sun12}.
Recently this phenomenon has been observed by imaging
classical light dynamics in a photorefractive crystal \cite{Sun12}.
In this Letter we show that the condensation of magnons in YIG \cite{Demokritov06,Demidov07,Dzyapko07,Demidov08a,Demidov08b,Demokritov08,Serga14,Clausen15a,Clausen15b} is another physical
realization of kinetic condensation of classical waves.

In order to calculate the nonequilibrium time evolution
of the pumped magnon gas in YIG, we model the  spin dynamics in YIG
by means of a stochastic Landau-Lifshitz-Gilbert \cite{Brown63} equation (LLG) with
non-Markovian damping due to the coupling to the thermal phonon bath.
The classical description is justified because 
 the relevant magnon energies are much smaller than
room temperature where the experiments are performed.
Moreover, the saturation magnetization 
of  YIG is rather large
so that the spins 
can be approximated by three-component classical vectors $\bd{S}_i ( t )$ 
of length $S \approx 14$
which are localized at the sites ${\bd{R}}_i$ of a cubic lattice \cite{Cherepanov93,Kreisel09}.
The coupling of the spins to the thermal phonon bath
gives rise to randomness and dissipation. Often these terms are taken into account
phenomenologically in  the LLG. Fortunately, for YIG 
the microscopic form of the dominant spin-phonon coupling is 
well known \cite{Gurevich96,Rueckriegel14}, so that we can derive the LLG 
microscopically by eliminating the
phonon coordinates from the equations of motion of the coupled spin-phonon 
system, as described in Refs.\cite{Rossi05,supplement}. In this way we arrive at the following stochastic LLG 
with non-Markovian damping,
\begin{eqnarray}
  \dot{\bd{S}}_i  ( t )  & = & \bd{S}_i ( t ) \times
 \Bigl[ {\bd{H}} ( t ) +  \bd{h}_i ( t ) + \sum_j \mathbb{K}_{ij} \bd{S}_j ( t ) 
 \Bigr]
 \nonumber
 \\
 &  - & 
 {\bd{S}}_i ( t ) \times  \int_0^t d t^\prime \sum_j 
 \mathbb{G}_{ij} ( t , t^\prime ) \dot{\bd{S}}_j ( t^\prime )   .
 \label{eq:LLG}
 \end{eqnarray}
Here the time-dependent external magnetic field  $\bd{H} ( t )$ is measured in units of energy
and $\mathbb{K}_{ij}$ is a matrix in the spin components with elements
$\mathbb{K}_{ij}^{\alpha \beta} = \delta^{\alpha \beta} J_{ij} + D^{\alpha \beta}_{ij}$,
where the exchange couplings $J_{ij}$ connect
nearest neighbor sites with strength $J$,
and $D^{\alpha \beta}_{ij}$ is the dipolar tensor \cite{Kreisel09,Rueckriegel14,supplement}.
The effect of the phonon bath on the spin dynamics manifests itself via the induced
magnetic field ${\bd{h}}_i ( t )$ and the
non-Markovian dissipation kernal $
 \mathbb{G}_{ij} ( t , t^\prime )$, which is a matrix in the spin components.
A microscopic derivation of these quantities starting from 
the coupled spin-phonon Hamiltonian for YIG is given in the Supplemental Material
\cite{supplement}. It turns out that the dissipation kernel
$\mathbb{G}_{ij} ( t , t^\prime )$ can be expressed in
terms of the properties of the phonon bath as follows:
 \begin{eqnarray}
   \mathbb{G}^{\alpha \beta}_{ij} ( t , t^\prime )  & = & \frac{1}{NS^4}
 \sum_{\mu \nu} B_{\alpha \mu} B_{ \beta \nu } S^{\mu}_i ( t ) S^\nu_j ( t^{\prime} )
\sum_{\bd{k} \lambda} e^{ i \bd{k} \cdot ( \bd{R}_i - \bd{R}_j ) }
 \nonumber
 \\
 & \times & 
  ( \bd{k}_{\alpha \mu} \cdot \bd{e}_{ \bd{k} \lambda} )
  ( \bd{k}_{ \beta \nu } \cdot \bd{e}_{ -\bd{k} \lambda} )
 \frac{
 \cos [ \omega_{\bd{k} \lambda} ( t - t^{\prime} )]}{ M \omega_{\bd{k}\lambda}^2 } ,
 \label{eq:Gdef}
 \end{eqnarray}
where $N$ is the number of lattice sites,
$\omega_{\bd{k} \lambda}$ is the dispersion of an acoustic phonon with
polarization vector $\bd{e}_{\bd{k} \lambda}$,  and $\lambda$ labels one longitudinal and two transverse phonon polarizations.
The effective ionic mass in the unit cell is denoted by $M$, and we have introduced the notation
 $\bd{k}_{\alpha \beta} = k_\alpha \bd{e}_{\beta} + k_{\beta} \bd{e}_{\alpha}$,
where $\bd{e}_{\alpha}$ are Cartesian unit vectors in direction $\alpha =x,y,z$.
Because for YIG  the magnetoelastic couplings $B_{\alpha \beta}$ and  all phonon
properties  are  known \cite{Gurevich96,Rueckriegel14},  we can explicitly calculate the dissipation kernel
$ \mathbb{G}^{\alpha \beta}_{ij} ( t , t^\prime )$.
The induced magnetic field
 $\bd{h}_i ( t )$ in Eq.~(\ref{eq:LLG}) is a linear functional
of the phonon coordinates, see the Supplemental Material \cite{supplement}. Assuming
that the phonons are in thermal equilibrium at temperature $T$, the 
 $\bd{h}_i ( t )$ are  Gaussian non-Markovian random processes.
While the average field $ \langle \bd{h}_i ( t ) \rangle = \bar{\bd{h}}_i ( t )$
leads only to a small 
correction to the external field $\bd{H} ( t )$ (which we neglect),
the fluctuating part $\delta \bd{h}_i ( t ) = \bd{h}_i ( t ) - \bar{\bd{h}}_i ( t )$
is crucial for the thermalization and the emergence of a magnon condensate.
As usual, the covariance of the induced magnetic field
is related to the damping kernel via the fluctuation-dissipation theorem,
$\langle \delta h_i^{\alpha} ( t ) \delta h_j^{\beta} ( t^{\prime} ) \rangle = T 
 \mathbb{G}_{ij}^{\alpha \beta} ( t , t^\prime )$.


The LLG (\ref{eq:LLG}) 
 describes the precession of the spins in an effective magnetic field, which includes 
the effect of spin-spin interactions as well as thermal fluctuations and damping caused by 
the random magnetic field obeying the 
fluctuation-dissipation theorem. However, unlike the
conventional stochastic LLG  the damping is described by a tensor
$\mathbb{G}_{ij} ( t , t^{\prime})$, 
which is  nonlocal in both space and time due to the nonlocal
spin-spin interactions induced by the exchange of phonons and
the memory effects of the thermal phonon bath.
This non-Markovian damping corresponds to a thermal noise field with a colored spectrum generated by the phonon dynamics. In contrast to the white noise in the conventional LLG 
this colored noise can also be expected to accurately describe the 
physical situation on small time scales and when the phonon and 
spin dynamics have  the same time scale \cite{Rossi05,Farias09b}.
Also note that the stochastic process with covariance (\ref{eq:Gdef}) 
is not ergodic due to the explicit  dependence of $\mathbb{G}_{ij} ( t , t^{\prime} )$
on the spin variables ${S}^{\mu}_i(t) S^{\nu}_j ( t^{\prime} )$.
Fortunately, in the experimentally used parallel pumping geometry
where the external magnetic field is of the form
 $ \bd{H} (t) = [{H}_0 + H_1 \cos(2\omega_{\rm p}t) ]\bd{e}_z $
(where $\omega_{\rm p}$ is the
pumping frequency and the static part $H_0$ of the magnetic field is much larger than
the amplitude $H_1$ of the oscillating part)
we can approximate $\mathbb{G}_{ij} ( t , t^{\prime} )$
by an ergodic process.
In this geometry 
the ground state of the spin system is a saturated 
ferromagnet magnetized in $z$-direction. 
Writing the spins as
$
 \bd{S}_i(t) = S [ \bd{e}_z + \bd{m}_i(t) ] $
it is then reasonable to 
expect that $|\bd{m}_i(t)| \ll 1$.
Given the fact that for YIG $S \approx 14$  is rather large,
we may substitute $\bd{S}_i ( t ) \rightarrow  S {\bd{e}}_z$
in Eq.~(\ref{eq:Gdef}). 
The damping kernel ${\mathbb{G}}_{ij} ( t , t^{\prime} )$
is then independent of the state of the spin system; i.e., it is ergodic.
Another benefit of this approximation is that 
${\mathbb{G}}_{ij} ( t , t^{\prime} )$ now depends only on the differences
$\bd{R}_i - \bd{R}_j$ and $t - t^{\prime}$ so that it is convenient to
work in momentum space and consider $\bd{m}_{\bd{k}} ( t ) = \sum_i e^{ - i \bd{k} \cdot 
 {\bd{R}}_i } \bd{m}_i ( t )$.
After this simplification, we can solve
our LLG numerically without further approximation. 
Technical details of the numerial solution can be found in the
Supplemental Material \cite{supplement}. In the rest of this Letter, we present our numerical results.

To make contact with the experiments \cite{Demokritov06,Demidov07,Dzyapko07,Demidov08a,Demidov08b,Demokritov08,Serga14,Clausen15a,Clausen15b} we have 
chosen the parameters for our simulation such that
they describe a typical experimental setup:
we consider a thin stripe of YIG with thickness
 $d = 6.7 \,{\rm \mu m} $ at temperature 
$T=300\,{\rm K}$ in a bias magnetic field
$ H_0 = 1710\, {\rm Oe} \times \mu $ and a pumping field $ H_1 = 0.03\, H_0 $,
where $\mu = 2 \mu_B$ is twice the Bohr magneton. 
The fields are assumed to be oriented parallel to the
longest axis of the stripe, which we identify with the $z$-axis.
Aligning the $x$-axis perpendicular to the stripe, the wave-vectors of the in-plane magnons
are of the form
$\bd{k}=k_y\bd{e}_y+k_z\bd{e}_z=|\bd{k}|\sin\theta_{\bd{k}}\bd{e}_y+|\bd{k}|\cos\theta_{\bd{k}}\bd{e}_z$ and the corresponding long-wavelength 
magnon dispersion in the absence of the pumping field is \cite{Kreisel09,Rueckriegel14}
\begin{eqnarray}
 E_{\bd{k}}& =& \left[ H_0 + \rho_s \bd{k}^2 + \Delta (1-f_{\bd{k}}) \sin^2\theta_{\bd{k}} \right]^{1/2} \nonumber\\ 
 &\times &\left[ H_0 + \rho_s \bd{k}^2 + \Delta f_{\bd{k}} \right]^{1/2} .
\end{eqnarray}
Here the energy  $\Delta = 4\pi \mu M_S$ 
is determined by  the saturation magnetization \cite{Gurevich96,Kreisel09,Tittmann73} 
$ 4 \pi M_S = 1750\, {\rm G}$, the spin stiffness is given by
 $\rho_s  =JSa^2  =  5.17 \times 10^{-9} \,{\rm Oe}\,{\rm cm}^2 \times \mu $, and the 
form factor is \cite{Kreisel09,Kalinikos86}
 $ f_{\bd{k}} = [ 1 - e^{-|\bd{k}|d} ]/ ({|\bd{k}|d}) $.
With the lattice spacing $a = 12.4 \,  {\rm \AA}$  we obtain for the effective spin
$S = M_S a^3 / \mu  \approx 14.2 $.
The magnon dispersion assumes then its minimal value
 $E_{\rm min}=2\pi\times 4.9\,{\rm GHz}$ for
$ \bd{k} = \pm k_{\rm \min} \bd{e}_z $ with $ k_{\rm min} = 4.999\times 10^4 \,{\rm cm}^{-1} $.
To fix the spin-phonon interaction and hence the damping kernel 
$\mathbb{G}_{ij} ( t , t^{\prime})$ we need the effective ionic mass \cite{Gilleo58,Gurevich96,Rueckriegel14}
$M =  5.17 \,{\rm g} \times a^3 /{\rm cm}^3$, the velocities
of the longitudinal and the transverse phonons,
$c_{\parallel} = 7.209\times 10^5 \, {\rm cm}/{\rm s}$, $
  c_{\bot} = 3.843\times 10^5 \, {\rm cm}/{\rm s}$, and the magnetoelastic coupling tensor,
which for the cubic lattice is of the form
$B_{\alpha\beta} = \delta_{\alpha\beta} B_\parallel + ( 1 - \delta_{\alpha\beta} ) B_\bot$.
For YIG, the relevant couplings are \cite{Gurevich96,Rueckriegel14}
  $B_\bot  \approx 2 B_{\parallel} = 6.96\times 10^6  \, {\rm erg} \times a^3 /{\rm cm}^3 $.
For the pumping frequency we choose  $\omega_{\rm p}=2\pi\times 7.046\,{\rm GHz}$, which is
slightly above the ferromagnetic resonance at $E_{\bd{k}=0}=2\pi\times 6.824\,{\rm GHz}$.
A convenient choice of the phonon polarization vectors 
$\bd{e}_{\bd{k} \lambda}$ appearing in Eq.~(\ref{eq:Gdef}) 
can be found in Ref.~[\onlinecite{Rueckriegel14}].
All parameters in Eqs.~(\ref{eq:LLG}) and (\ref{eq:Gdef}) are now fixed so that
the results of our numerical simulations do not contain any adjustable parameters.

\begin{figure}[t]
\includegraphics[width=40mm]{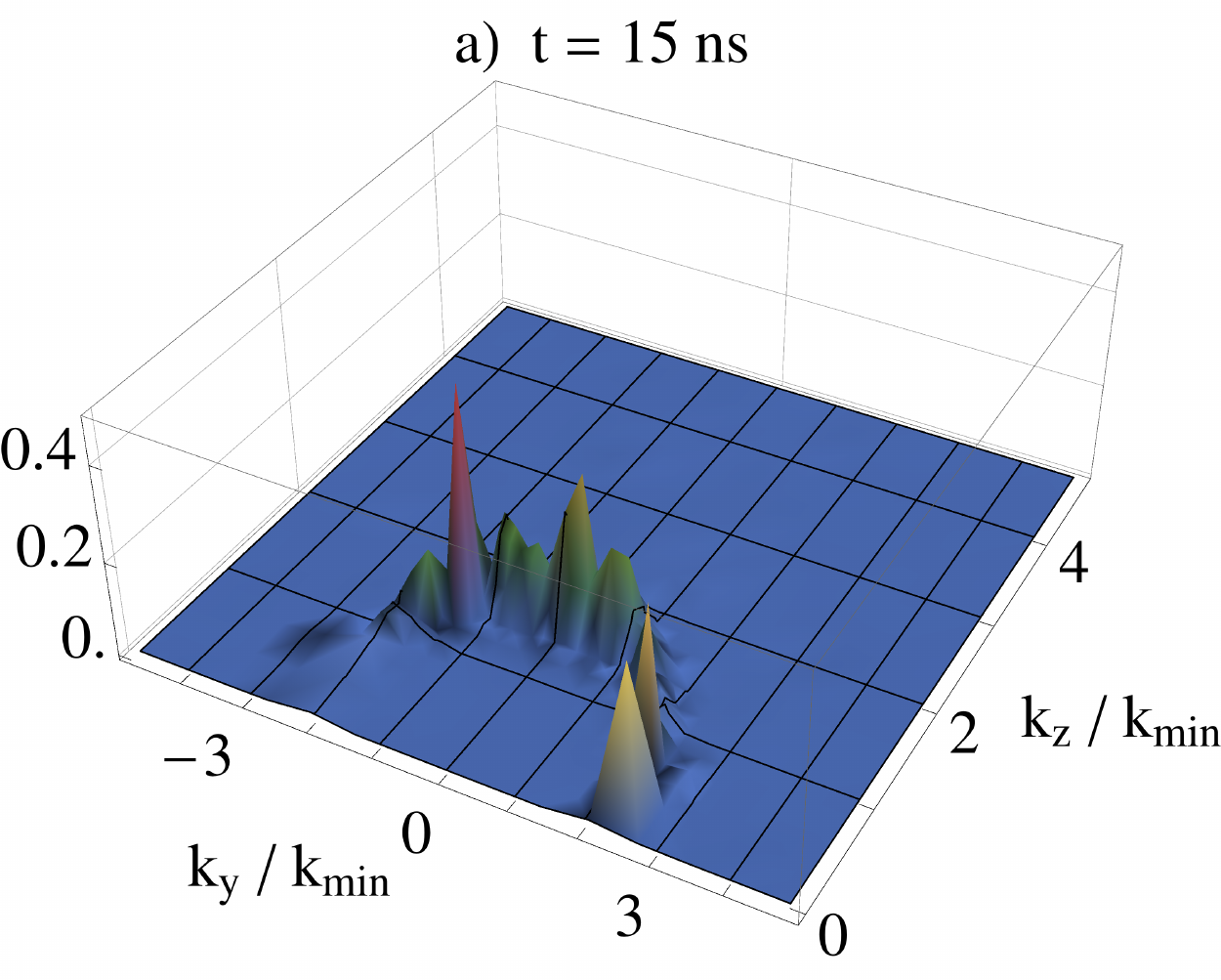}
\includegraphics[width=40mm]{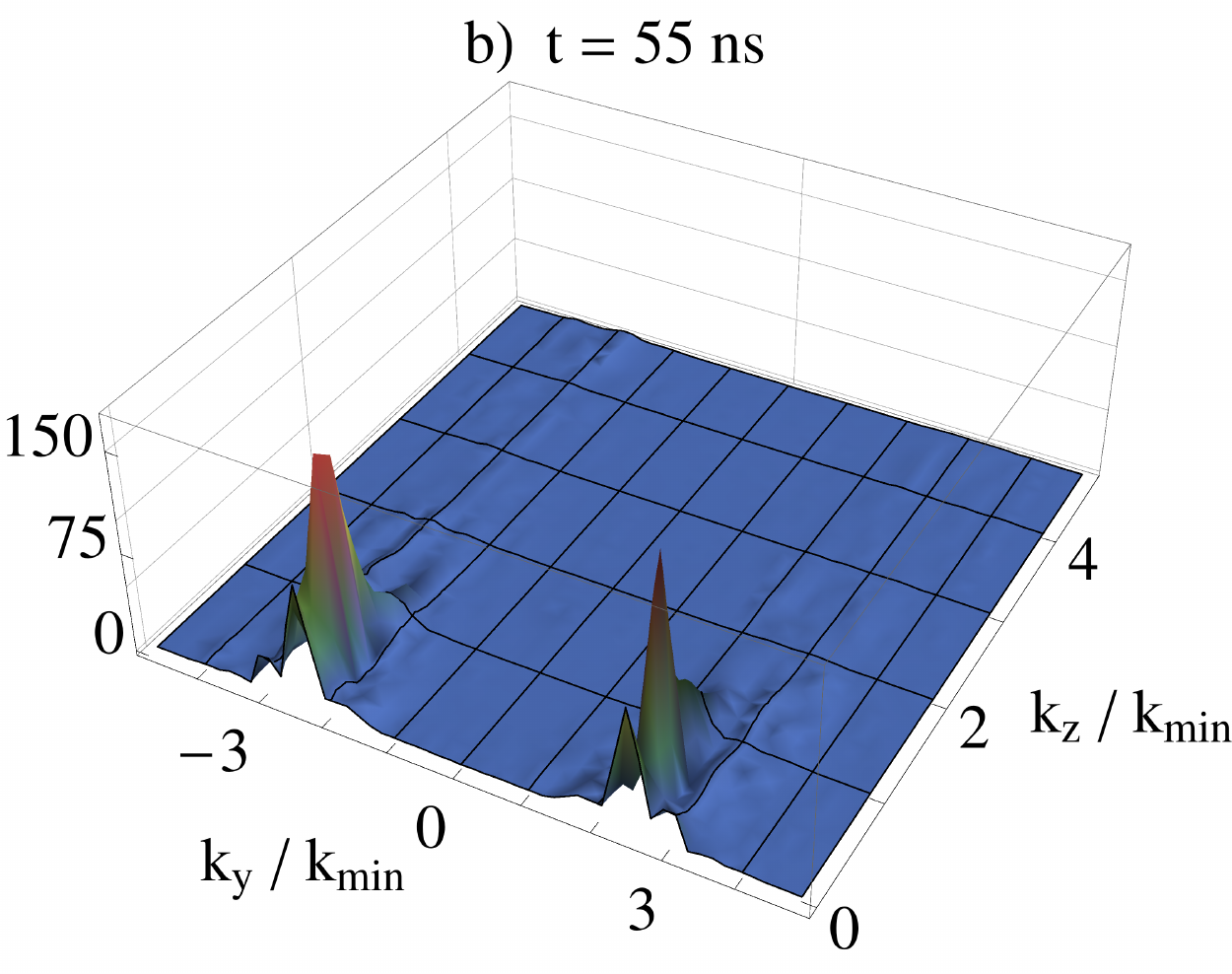}
  \\
\includegraphics[width=40mm]{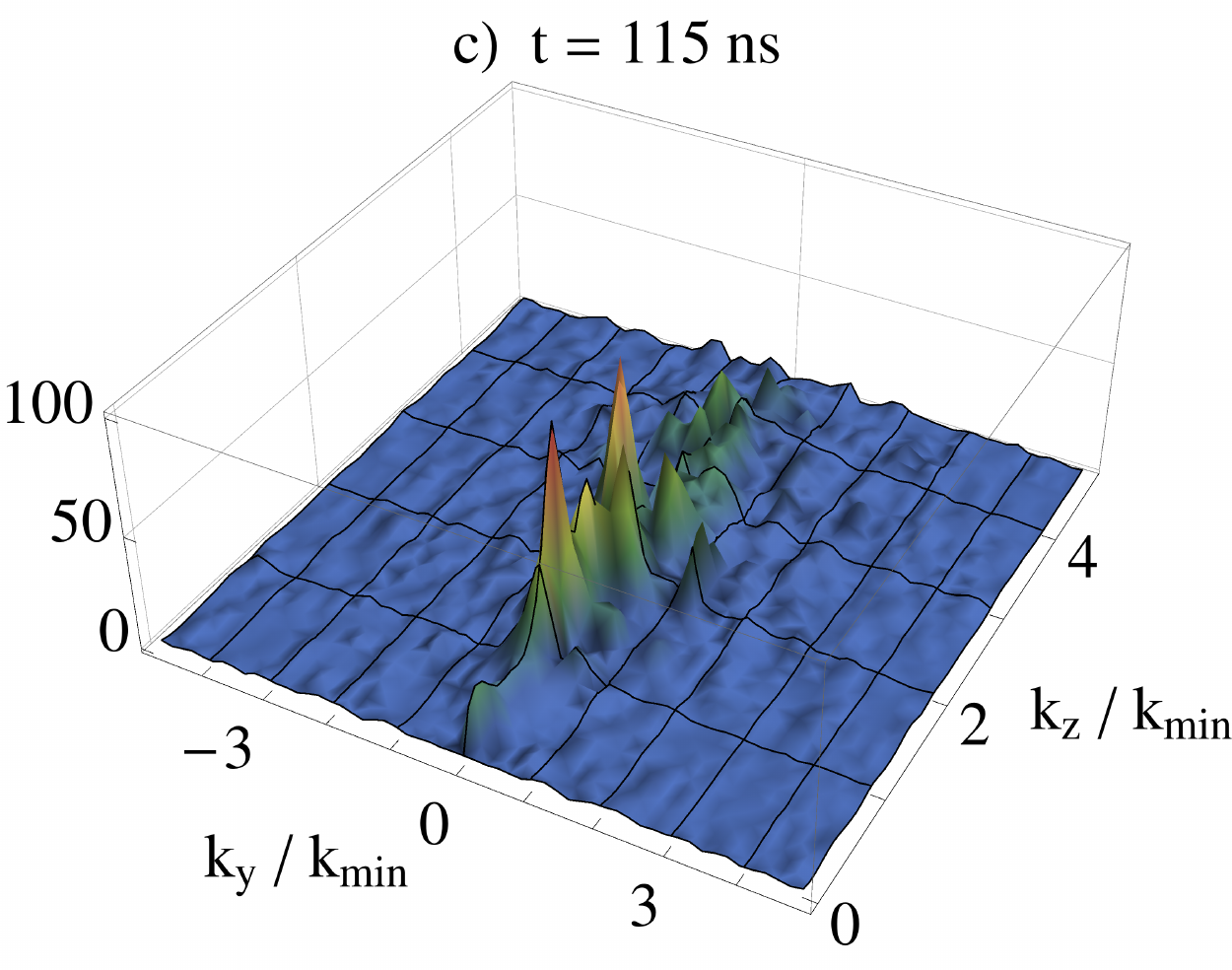}
\includegraphics[width=40mm]{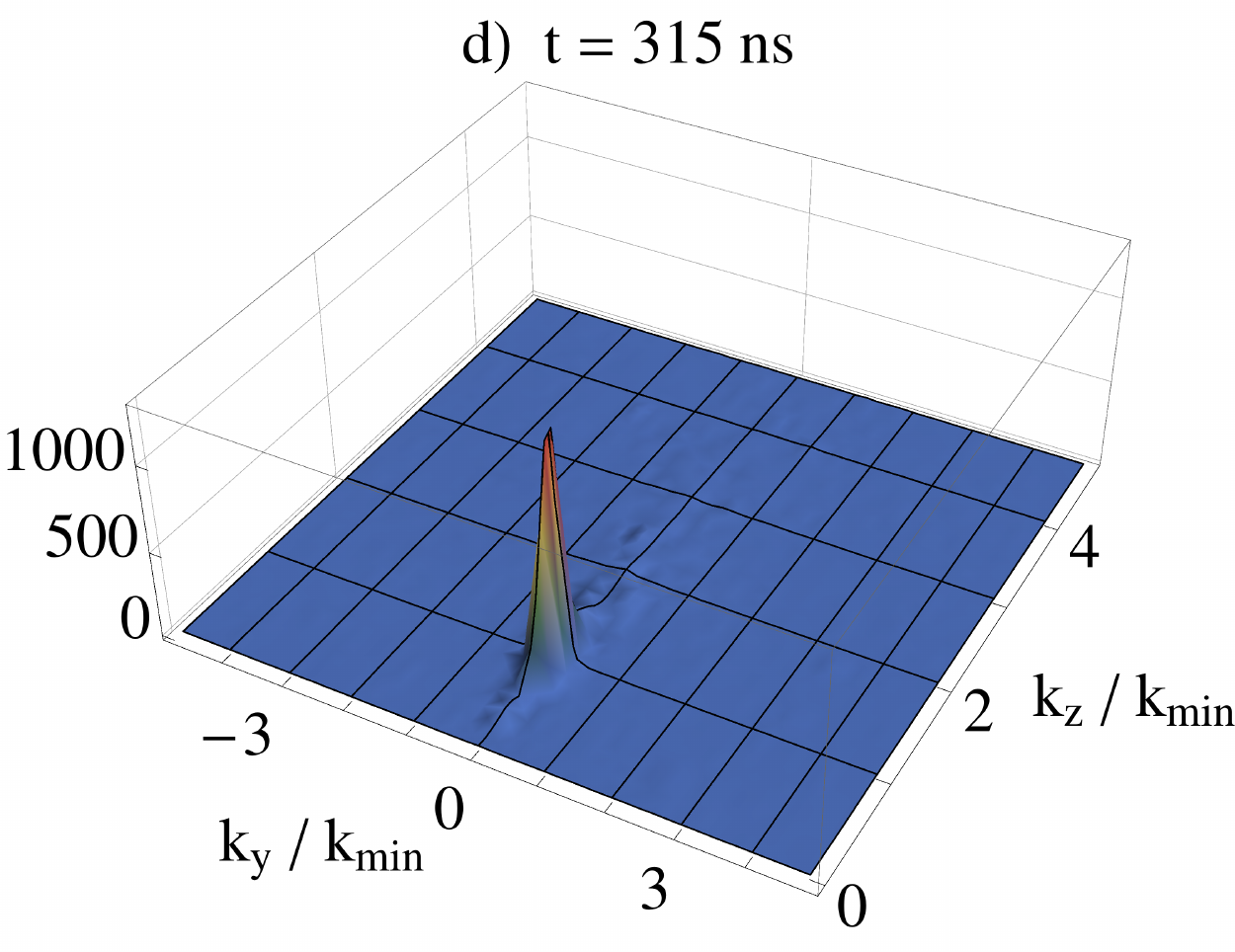}
  \caption{%
(Color online).
Snapshots of the stochastic time evolution of the magnon distribution $n_{\bd{k}} ( t ) / n_{\bd{k}}^{\rm th}$ relative to the thermal distribution $ n_{\bd{k}}^{\rm th} $, 
showing (a) the magnetoelastic hybridization, (b) the parametric instability, 
(c) the redistribution of magnons in momentum space, and (d) 
the quasiequilibrium with the magnon condensate.
}
  \label{fig:evolution}
\end{figure}

\begin{figure}[t]
\includegraphics[width=42mm]{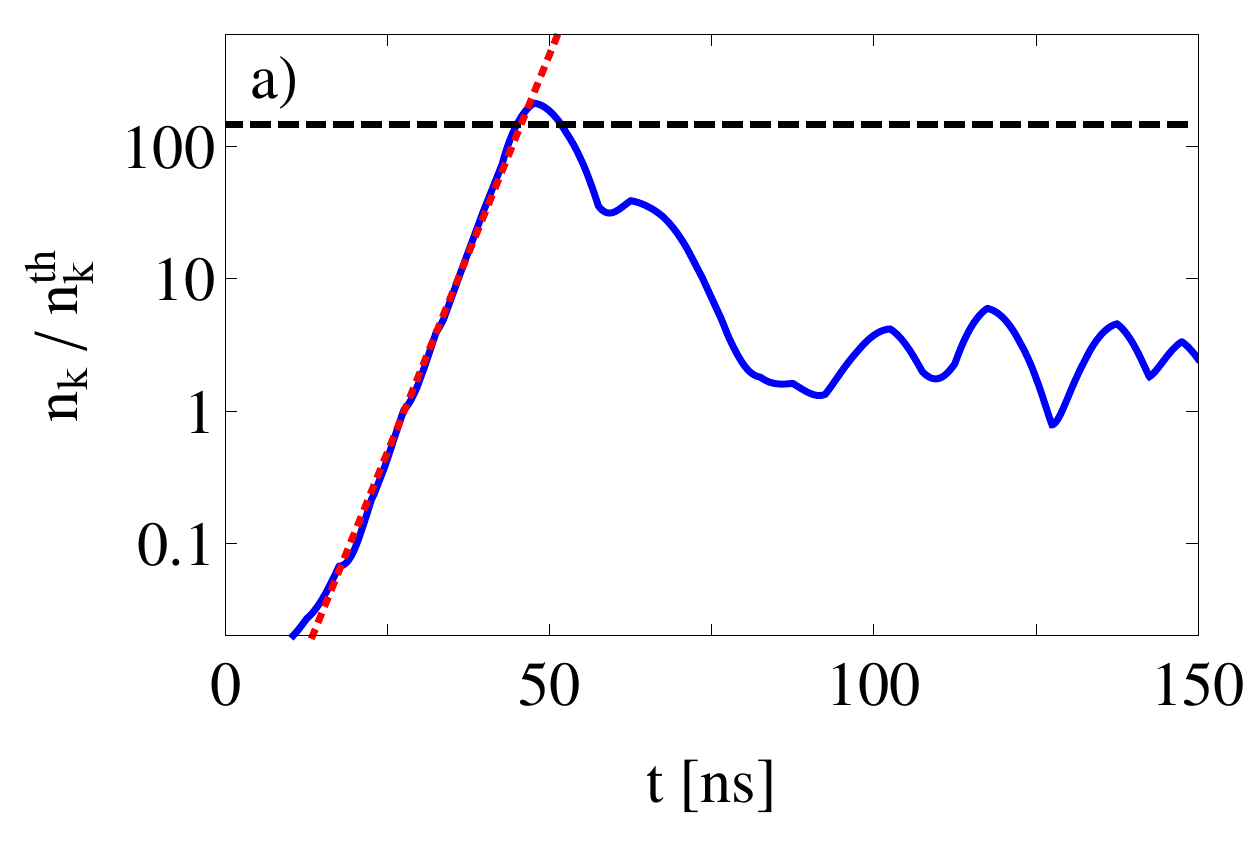}
\includegraphics[width=42mm]{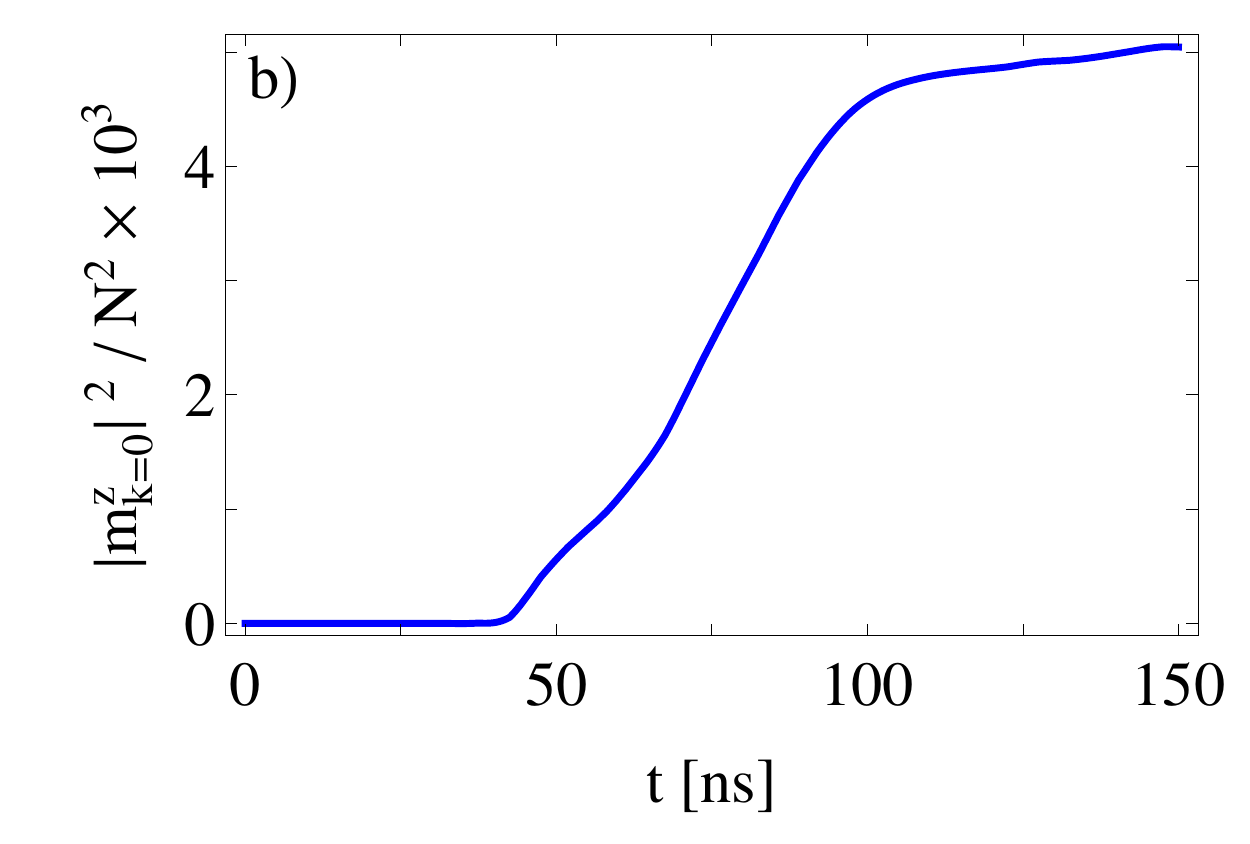}
  \caption{%
(Color online).
(a)
Logarithmic plot of the parametric resonance. 
The blue solid line shows $n_{\bd{k}} ( t )/ n_{\bd{k}}^{\rm th}$
for the parametrically unstable magnon with
momentum $\bd{k}/k_{\rm min} = -2.4 \, \bd{e}_y + 0.2 \, \bd{e}_z $.
For comparison, the red dotted line is
the exponential growth  predicted by linear spin wave theory
as discussed in the main text.
Note that in our nonlinear theory the growth saturates at some finite value close to 
the steady state 
predicted by S-theory (black dashed line).
At longer times, the magnon distribution relaxes 
to a quasiequilibrium value.
(b) Plot of the stochastic time evolution of the uniform mode
$|m^z_{\bd{k}=0}|^2/N^2$, which 
grows rapidly as soon as
the growth of parametrically unstable modes saturates 
at $t\approx 50\,{ \rm ns}$, and saturates itself when the parametric magnons thermalize to quasiequilibrium around $t\approx 100\,{\rm ns}$.
}
  \label{fig:PI}
\end{figure}

In the following we present plots of the square of the 
 transverse spin component
 $n_{\bd{k}}(t ) =  |m_{\bd{k}}^x(t)|^2 + | m_{\bd{k}}^y(t) |^2$.
The average $\langle  n_{\bd{k}}(t ) \rangle$   is directly proportional to the 
magnon number of spin wave theory \cite{Gurevich96}.
In particular, in thermal equilibrium
$\langle  n_{\bd{k}}(t ) \rangle  = n_{\bd k}^{\rm th} \propto T / E_{\bd{k}}$ is proportional
to the classical Rayleigh-Jeans distribution.
In  Fig. \ref{fig:evolution} we illustrate the main stages
of the time evolution of  $n_{\bd{k}}(t )$, assuming that the system
is prepared in a fully polarized ferromagnetic state when the pumping field is switched on.
%
%
The time evolution can be characterized by four distinct regimes:
The population of the magnon spectrum begins on the energy surface where 
magnetic and elastic modes hybridize, as shown in Fig.~\ref{fig:evolution}~(a).
Magnon-phonon hybridization in YIG has recently been discussed in Ref.~[\onlinecite{Rueckriegel14}].
Note that our non-Markovian description is crucial for obtaining the magnon-phonon hybridization \cite{Maluckov13}. 
At the next stage shown in Fig.~\ref{fig:evolution}~(b)
the growth of the magnon distribution
is dominated by the parametric instability of magnons
with energies $E_{\bd{k}}$ in the vicinity of the pumping frequency $\omega_{\rm p}$.
An example of a stochastic trajectory of the parametrically pumped  magnon with
 $E_{\bd{k}} = \omega_{\rm p}$
is displayed in Fig.~\ref{fig:PI}~(a). 
%
%
%
%
Note that linear spin wave theory predicts that 
in the regime of parametric instability the envelope of
the magnon occupation increases exponentially \cite{Gurevich96,supplement},
$n_{\bd{k}} ( t ) \propto e^{ 2 \alpha_{\bd{k}} t }$ with
$\alpha_{\bd{k}}^2 =  |  V_{\bd{k}} H_1 |^2 - | E_{\bd{k}} - \omega_{\rm p}  |^2$.
Here $V_{\bd{k}}$
is proportional to the so-called ellipticity
of the spin waves and can be expressed in terms of
the Fourier transform $D^{\alpha \beta}_{\bd{k}}$ of the dipolar tensor as
$ V_{\bd{k}}
 =  S ( D_{\bd{k}}^{xx} - D_{\bd{k}}^{yy} )/( 4 E_{\bd{k}} )$.
Because the growth coefficient $\alpha_{\bd{k}}$ depends 
on the ellipticity, the parametric pumping is most effective
for  elliptic magnons close to $k_z=0$.

\begin{figure}[t]
\includegraphics[width=80mm]{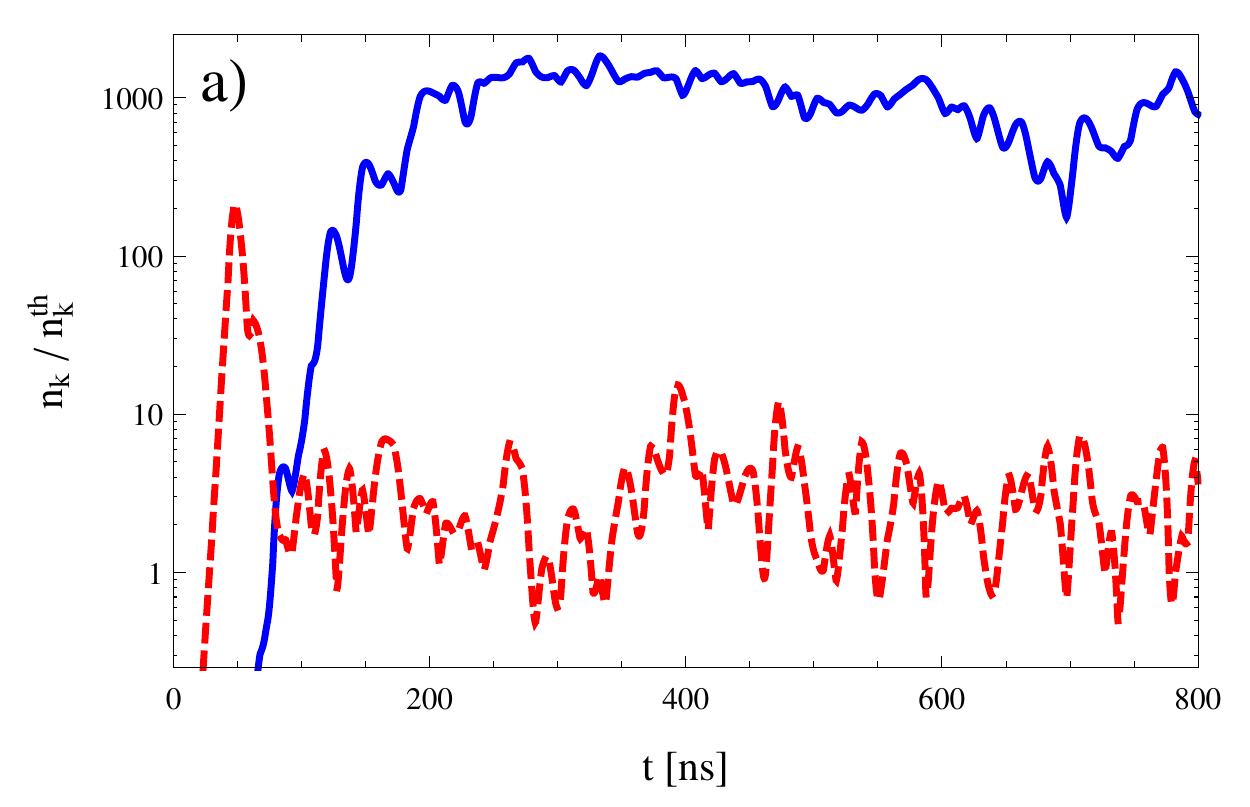}
 \\
\includegraphics[width=80mm]{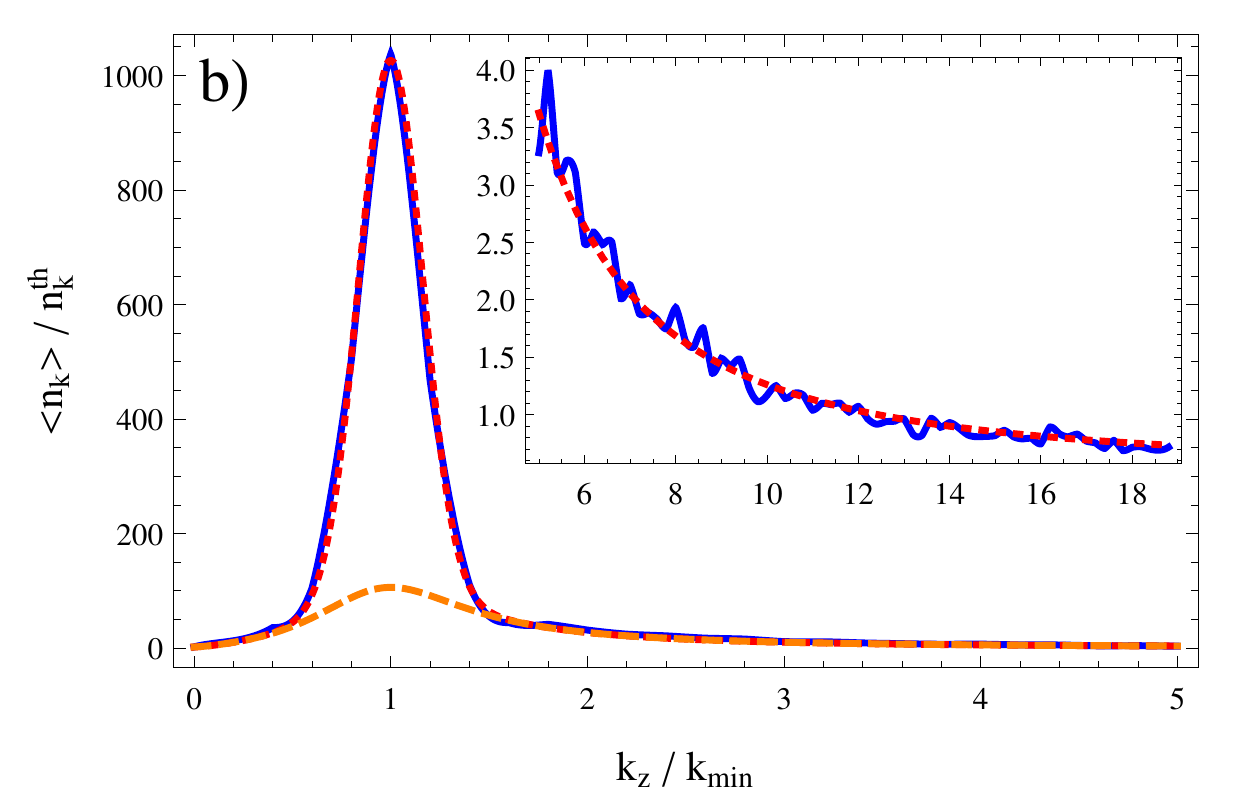}
  \caption{%
(Color online).
(a)
Logarithmic plot of the stochastic time evolution of $n_{\bd{k}} (t)/ n_{\bd{k}}^{\rm th}$
of the condensing mode $\bd{k}=k_{\rm min}\bd{e}_z$ (blue solid line), 
and of the parametrically unstable mode with
momentum $\bd{k}/k_{\rm min} = -2.4 \, \bd{e}_y + 0.2 \, \bd{e}_z $ (red dashed line).
Note that the growth of the condensate begins as soon as the parametrically excited magnons 
have relaxed to quasiequilibrium.
(b)
Time-averaged quasiequilibrium distribution function 
$\langle n_{\bd{k}} ( t ) \rangle / n_{\bd{k}}^{\rm th} $
(blue solid line) on the axis $k_y=0$. 
The red dotted line shows the fit given in Eq. (\ref{eq:quasiEq}) with $\mu_{\rm eff}= 0.995\, E_{\rm min}$, 
while the orange dashed line is the thermal part alone.
The condensate contribution was modeled as a Gaussian centered 
at $\bd{k}=k_{\rm min} \bd{e}_z$ with standard deviations 
$\sigma_z=0.16\,k_{\rm min}$ and $\sigma_y=0.1\,k_{\rm min}$, which are both smaller than the lattice spacing $\Delta k=0.2\,k_{\rm min}$.
The inset shows a magnified view of the fit away from the condensate.
}
  \label{fig:BEC}
\end{figure}

Because of interaction effects (which are nonperturbatively taken into account
in our approach) the exponential growth
of  the parametrically pumped magnons  stops
after approximately $40 \,{\rm ns}$. From Fig.~\ref{fig:PI}~(b) we see that  
at this point the uniform longitudinal mode
 $m^z_{\bd{k}=0}$ starts to be populated, 
although the occupation always remains small.
S-theory provides an estimate for the saturation value of the parametric magnons \cite{Rezende09,Lvov94,Zakharov70}
 $ n_{\bd{k}}^S / n_{\bd{k}}^{\rm th} \approx
 N S E_{\bd{k}} ( | V_{\bd{k}} H_1 | - 
| E_{\bd{k}} - \omega_{\rm p} | )  /(8 T \Delta)  $,
which is shown as a black dashed line in Fig.~\ref{fig:PI}~(a) 
and agrees reasonably well with our simulation.
In S-theory, this corresponds to a steady state, but our
simulation reveals that 
at $t\approx 50\,{\rm ns}$
the parametric magnons begin to relax to a quasiequilibrium value although the pumping field is still present. 
This is the third stage of the time evolution shown in Fig.~\ref{fig:evolution}~(c), where
excited magnons generated via the parametric instability are scattered 
to the states with $k_y=0$. Eventually, in the fourth stage
shown in Fig.~\ref{fig:evolution}~(d), the magnons 
accumulate at 
$k_z=k_{\rm min}$ to form the Rayleigh-Jeans condensate. 
The stochastic time evolution of the condensed mode is shown in Fig.~\ref{fig:BEC}~(a), 
with the primary parametric modes for comparison.
The formation of the condensate is completed at $t\approx 200\, {\rm ns}$;
at later times, there are only small oscillations around a stable quasiequilibrium state
for all modes. This allows us to obtain thermal averages by averaging over time. 
Choosing the interval between 
 $200$ and $ 800\,{\rm ns}$ for the average we obtain
the quasiequilibrium distribution 
\begin{equation}
 \langle n_{\bd{k}} ( t ) \rangle / n_{\bd{k}}^{\rm th} \propto
 E_{\bd{k}} /( E_{\bd{k}} - \mu_{\rm eff}  ) + n_{\rm c} 
 \delta( \bd{k}\mp k_{\rm min}\bd{e}_z) .
 \label{eq:quasiEq}
\end{equation}
The first part corresponds to a rescaled Rayleigh-Jeans  
distribution with effective chemical potential
$\mu_{\rm eff}\approx E_{\rm min}$, which is valid for all modes with the exception of the
modes at the dispersion minima, for which one has to add 
delta-like contributions with weight $n_{\rm c} $ 
and width comparable to the resolution $\Delta k$ of the momentum grid.
This proves that the classical spin dynamics  indeed gives rise
to Rayleigh-Jeans condensation of magnons at the minima of the energy dispersion.
A plot of the time averaged distribution function as well as of the fitting 
function (\ref{eq:quasiEq}) is displayed in Fig.~\ref{fig:BEC}~(b). 
Let us emphasize that the retardation of the dissipation kernel in Eq.~(\ref{eq:Gdef}) is
not essential for the appearance of the condensate.
However, if we use a phenomenological dissipation kernel with a shorter correlation time
a condensate emerges only for unrealistic values of the magnon-phonon coupling and the pumping field.

\begin{figure}[t]
\includegraphics[width=42mm]{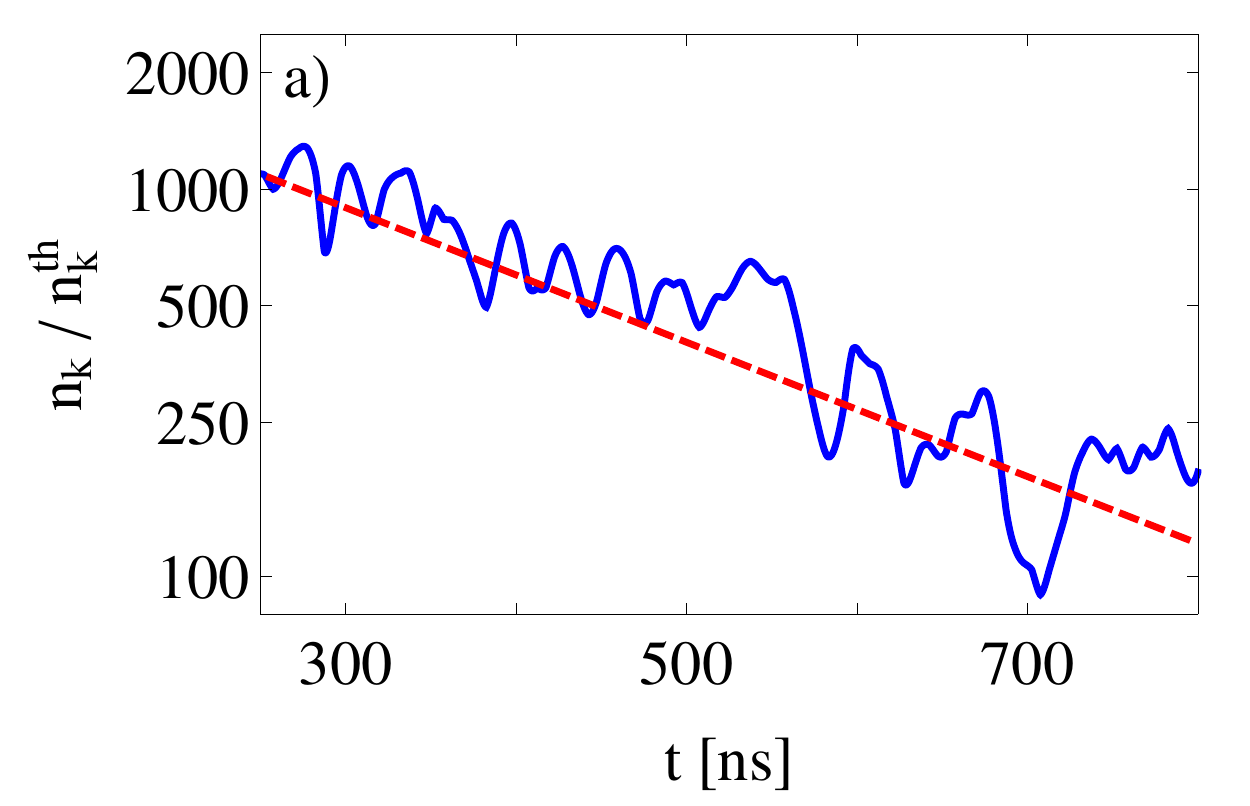}
\includegraphics[width=42mm]{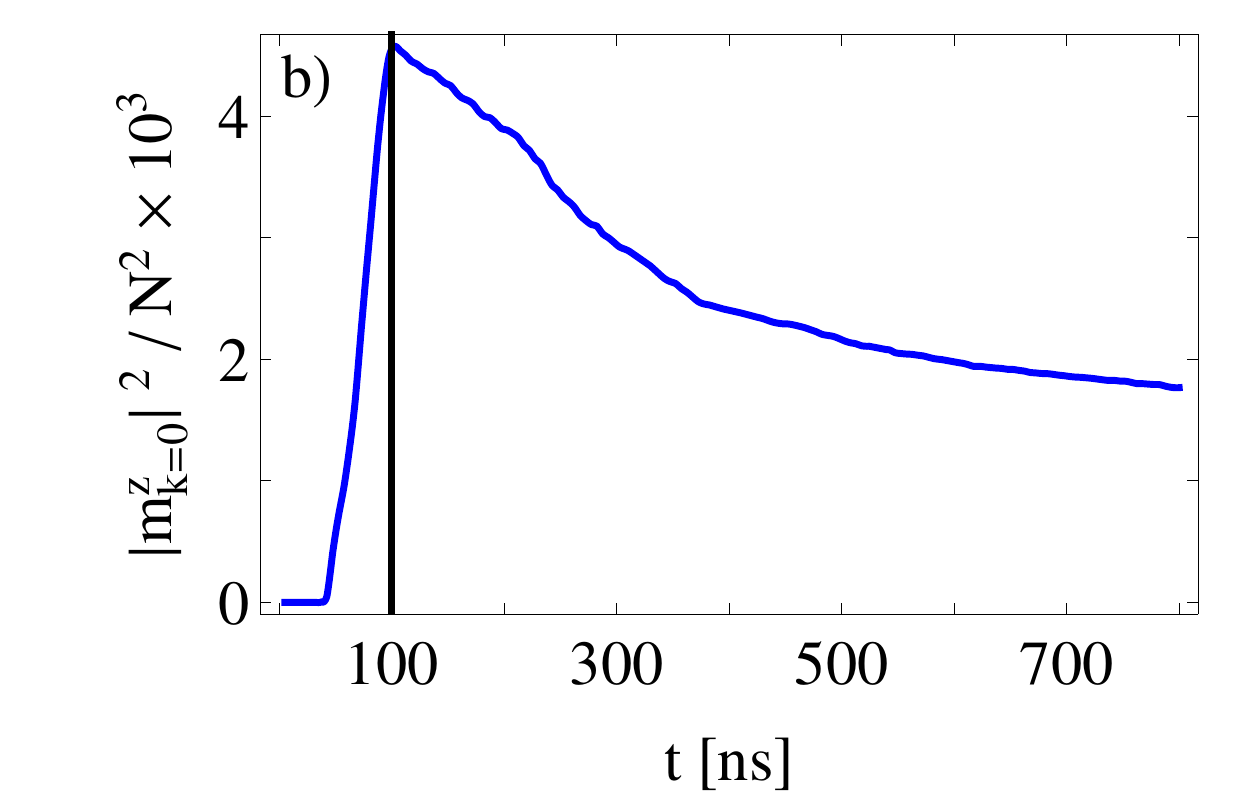}
 \caption{%
(Color online).
(a) Logarithmic plot of the decay of the condensate. 
The blue line denotes $n_{\bd{k}} (t)/ n_{\bd{k}}^{\rm th}$ at  $\bd{k}=k_{\rm min}\bd{e}_z$, whereas 
the red dashed line is an exponential fit with the condensate lifetime $\tau_{\rm c} \approx 250\,{\rm ns} $.
(b) Plot of the stochastic time evolution of the uniform mode $|m^z_{\bd{k}=0}|^2/N^2$. As soon as the pumping is turned off at $t=100\,{\rm ns}$
the uniform mode starts to decay.
}
 \label{fig:BECdecay2}
 \end{figure}

Finally, we show in Fig.~\ref{fig:BECdecay2} the time evolution of the system
for the case that the pumping field is switched off at $t=100\,{\rm ns}$. 
The condensate still forms in this regime, in agreement with the experiments \cite{Demidov08a,Demidov08b,Demokritov08,Clausen15a}.
However, at $t \approx 250\,{\rm ns}$ it begins to decay, whereas the uniform mode decays as soon as the pumping field is turned off.
Moreover,
from the left panel it is clear 
that the decay of the
magnon condensate is to a good approximation exponential,
with  characteristic decay time
$\tau_{\rm c} \approx 250 \; {\rm ns}$, in excellent agreement with the experiments \cite{Demokritov08,Clausen15b}.


%
%
%

\


In summary, we have derived and solved 
a non-Markovian stochastic LLG  which describes all stages of the  nonequilibrium time
evolution  of the pumped magnon gas in YIG, including the
magnon-phonon hybridization at early stages, followed by the
parametric resonance regime and the emergence of a magnon condensate, and 
the eventual decay of the condensate and the thermalization of the magnon distribution
after the pumping has been switched off.
Such a complete description of all stages leading to the quasiequilibrium condensation of magnons in YIG has not been achieved before. 
Our results are in quantitative agreement with experiments and prove that
the emergence of a condensate in the pumped magnon gas in YIG
is a purely classical kinetic condensation phenomenon, similar to
the condensation of classical light which has been observed in photorefractive 
crystals~\cite{Sun12}.

We thank A. A. Serga and D. A. Bozhko for discussions and the 
DFG for financial support via SFB/TRR 49.

\pagebreak


\setcounter{equation}{0}
\setcounter{figure}{0}
\setcounter{table}{0}
\setcounter{page}{1}
\makeatletter
\renewcommand{\theequation}{S\arabic{equation}}
\renewcommand{\thefigure}{S\arabic{figure}}
\renewcommand{\bibnumfmt}[1]{[S#1]}
\renewcommand{\citenumfont}[1]{S#1}


\section{Supplemental Material}

The following supplementary material contains technical details
of the calculations presented in the main text and some additional derivations.

\appendix

\section{Stochastic Landau-Lifshitz-Gilbert equation 
from microscopic spin-phonon dynamics   }

Here we derive the stochastic non-Markovian Landau-Lifshitz-Gilbert equation
given in Eq.~($1$) of the main text starting from the quantum mechanical
Hamiltonian for the coupled spin-phonon system in YIG.
Our derivation follows the work by Rossi, Heinonen, and MacDonald \cite{Rossi05s}.
The established Hamiltonian of the relevant
spin- and lattice degrees of freedom in YIG  consists of three parts \cite{Rueckriegel14s},
\begin{equation}
 {\cal{H}} ( t ) = {\cal{H}}_{\rm m} ( t ) + {\cal{H}}_{\rm e} + {\cal{H}}_{\rm me},
 \label{eq:H3}
 \end{equation}
where  the magnetic part ${\cal{H}}_{\rm m} ( t )$ depends only on the spin degrees of freedom, 
 ${\cal{H}}_{\rm e} ( t )$ is the elastic (phonon) part,
and ${\cal{H}}_{\rm me}$ is the magnetoelastic coupling.
Assuming that the system is exposed to a time-dependent external 
magnetic field $\bd{H} ( t )$, 
the magnetic part of the Hamiltonian is
 \begin{eqnarray} 
 {\cal{H}}_{\rm m} ( t )  &=& -\frac{1}{2} \sum_{ij} \sum_{\alpha \beta}
 [  \delta^{\alpha \beta}   J_{ij}  + D_{ij}^{\alpha \beta} ]
{{S}}^{\alpha}_i  {{S}}^{\beta}_j 
 \nonumber
 \\
& &
 -  \sum_i  \bd{H} ( t ) \cdot \bd{S}_i ,
 \label{eq:Hspin}
\end{eqnarray} 
where $\bd{S}_i$ are quantum mechanical spin operators
localized at sites $\bd{R}_i$ of a cubic lattice, the
$J_{ij}$ are the ferromagnetic exchange couplings
connecting nearest neighbors on a cubic lattice,
and the matrix elements
$ D_{ij}^{\alpha \beta}$ of the dipolar tensor
are ($\mu_B$ is the Bohr magneton),
 \begin{equation}
 D_{ij}^{\alpha \beta}  = (1- \delta_{ij})\frac{(2 \mu_B)^2}{|\bd{R}_{ij}|^3} \left[ 3\hat R_{ij}^\alpha \hat R_{ij}^\beta - \delta^{\alpha \beta}  \right],
\label{eq:dipdef}
\end{equation}  
where
${\bd R}_{ij}={\bd R}_i - {\bd R}_j$  and $\hat {\bd R}_{ij} = {\bd R}_{ij}/|{\bd R}_{ij}|$.
Defining the tensor $\mathbb{K}_{ij}$ with matrix elements
$\mathbb{K}_{ij}^{\alpha \beta} =  \delta^{\alpha \beta}   J_{ij}  + D_{ij}^{\alpha \beta}$,
the spin Hamiltonian (\ref{eq:Hspin}) can alternatively be written as
 \begin{equation}
 {\cal{H}}_{\rm m} ( t )   = -\frac{1}{2} \sum_{ij} 
 {\bd{S}}_i  \mathbb{K}_{ij} {\bd{S}}_j 
 -  \sum_i  \bd{H} ( t ) \cdot \bd{S}_i .
 \label{eq:Hspin2}
\end{equation} 
The elastic part of our Hamiltonian is given by
\begin{equation}
   {\cal{H}}_{\rm e} =  \frac{1}{N} \sum_{\bd{k} \lambda}
 \left[ \frac{ P_{ - \bd{k} \lambda} P_{\bd{k} \lambda}}{2 M}
 + \frac{M}{2} \omega_{\bd{k} \lambda}^2 X_{ - \bd{k} \lambda}
 X_{\bd{k} \lambda} \right],
\end{equation}
where $M$ is the effective ionic mass of a unit cell and
$\omega_{\bd{k} \lambda} = c_{\lambda} | \bd{k} |$
is the dispersion of acoustic phonons with polarization $\lambda$
and velocity $c_{\lambda}$.
The canonical momenta
$P_{\bd{k} \lambda}$ and coordinates $X_{\bd{k} \lambda}$ associated with the normal modes satisfy the
commulation relations $[ X_{\bd{k} \lambda} , P_{\bd{k}^{\prime} \lambda^{\prime}} ] = i
N\delta_{ \bd{k} , - \bd{k}^{\prime}} \delta_{\lambda  \lambda^{\prime}}$.
With our normalization of the Fourier expansion,
the lattice displacement
 $\bd{X}(\bd{R}_i)$ at lattice site $\bd{R}_i$ is given by
\begin{equation}
 \bd{X}(\bd{R}_i)= \frac{1}{N} \sum_{\bd{k} \lambda} e^{i\bd{k}\cdot\bd{R}_i} X_{\bd{k} \lambda} \bd{e}_{\bd{k} \lambda} ,
\end{equation}
where $N$ is the number of lattice sites and
$\bd{e}_{\bd{k} \lambda}$ are the polarization vectors
associated with the normal mode with momentum $\bd{k}$ and
polarization $\lambda$.
Finally, the 
magnetoelastic coupling in YIG is to leading order in a gradient expansion
given by \cite{Gurevich96s,Rueckriegel14s}
 \begin{equation}
 {\cal{H}}_{\rm me} = \frac{1}{S^2} \sum_i \sum_{\alpha \beta} B_{\alpha \beta} 
 S^{\alpha}_i S^\beta_{i} X_i^{\alpha \beta},
 \end{equation}
where the strain tensor is defined in terms of the components
$X_{\alpha} ( \bd{r} )$ of the phonon displacements in direction $\bd{e}_{\alpha}$,
 \begin{equation}
 X_i^{\alpha \beta } = \frac{1}{2} 
 \left[ \frac{ \partial X_{\alpha} ( \bd{r} ) }{\partial r_{\beta} }
 +  \frac{ \partial X_{\beta} ( \bd{r} ) }{\partial r_{\alpha} }
 \right]_{\bd{r} = \bd{R}_i }.
 \end{equation}
For YIG the magnetoelastic coupling tensor is of the form \cite{Gurevich96s,Rueckriegel14s}
 \begin{equation}
 B_{\alpha \beta } = \delta_{\alpha \beta } B_{\parallel} +
(1 - \delta_{\alpha \beta} ) B_{\bot} .
 \end{equation}

From the Hamiltonian (\ref{eq:H3}) it is straightforward to obtain the
quantum mechanical equation of motion
for the spin operators,
 \begin{eqnarray}
 \dot{\bd{S}}_i (t )  & = & \bd{S}_i ( t ) \times
 \Big[ {\bd{H}} ( t ) + \sum_j \mathbb{K}_{ij} \bd{S}_j ( t ) 
 \Big]
 \nonumber
 \\
 &+ & \frac{1}{2} \left[ \bd{S}_i ( t ) \times {\bd{F}}_i ( t )
 -   {\bd{F}}_i ( t ) \times \bd{S}_i ( t ) \right],
 \label{eq:Heisenbergpho}
 \hspace{7mm}
 \end{eqnarray}
where the vector operators $ {\bd{F}}_i ( t )$ describe the effect of the
magnetoelastic coupling on the spin-dynamics,
 \begin{equation}
  {\bd{F}}_i (t )
 = - \frac{2}{S^2} \sum_{\alpha \beta} B_{\alpha \beta} \bd{e}_{\alpha} X_i^{\alpha \beta} (t )
 S^{\beta}_i (t).
 \end{equation}
The equation of motion (\ref{eq:Heisenbergpho}) should be complemented
by the equations of motion for the normal modes of the
phonons,
 \begin{eqnarray}
 \dot{X}_{\bd{k} \lambda} & = & \frac{ P_{\bd{k} \lambda}}{M},
 \label{eq:XEoM}
 \\
  \dot{P}_{\bd{k} \lambda} & = & - M \omega_{\bd{k} \lambda}^2  X_{\bd{k} \lambda}
 +  M A_{\bd{k} \lambda}  (t),
 \end{eqnarray}
where
 \begin{eqnarray}
  A_{\bd{k} \lambda}  (t) & = & \frac{i}{2 M S^2 } \sum_i \sum_{\alpha \beta}
  e^{ - i \bd{k} \cdot {\bd{R}}_i } B_{\alpha \beta}
 \nonumber
 \\
 & & \times
 {\bd{k}}_{\alpha \beta} \cdot {\bd{e}}_{ - \bd{k} \lambda} S_i^{\alpha} (t )
 S_i^{\beta} ( t ).
 \end{eqnarray}
Here $\bd{k}_{\alpha \beta} = k_{\alpha} \bd{e}_{\beta} + k_{\beta} \bd{e}_{\alpha}$. 
Taking an additional time derivative of  Eq.~(\ref{eq:XEoM})
we obtain
 \begin{equation}
 ( \partial_t^2 + \omega_{\bd{k} \lambda}^2 ) X_{\bd{k} \lambda} =
 A_{\bd{k} \lambda} (t ) ,
 \end{equation}
which has the general solution
 \begin{eqnarray}
 X_{\bd{k} \lambda} (t) & = & X_{\bd{k} \lambda} ( 0 ) \cos ( \omega_{\bd{k} \lambda} t )
 + \frac{ P_{\bd{k} \lambda} (0)}{M \omega_{\bd{k} \lambda} } 
 \sin ( \omega_{\bd{k} \lambda} t )
 \nonumber
 \\
 &  & + \int_0^t d t^{\prime} \frac{ \sin [ \omega_{\bd{k} \lambda} (t - t^{\prime}) ]}{
 \omega_{\bd{k} \lambda} } A_{\bd{k} \lambda} ( t^{\prime} )
 \nonumber
 \\
 & = &  \tilde{X}_{\bd{k} \lambda} ( 0 ) \cos ( \omega_{\bd{k} \lambda} t )
 + \frac{ P_{\bd{k} \lambda} (0)}{M \omega_{\bd{k} \lambda} } 
 \sin ( \omega_{\bd{k} \lambda} t )
 \nonumber
 \\
 &  & + \frac{ A_{\bd{k} \lambda} ( t ) }{ \omega_{\bd{k} \lambda}^2 }
  - \int_0^t d t^{\prime} \frac{ \cos [ \omega_{\bd{k} \lambda} (t - t^{\prime}) ]}{
 \omega_{\bd{k} \lambda}^2 } \frac{ d A_{\bd{k} \lambda} ( t^{\prime} )}{  d t^{\prime} }.
 \nonumber
 \\
 & & 
 \end{eqnarray}
In the last line, we have integrated by parts and introduced the notation
 \begin{equation}
   \tilde{X}_{\bd{k} \lambda} ( 0 ) =  {X}_{\bd{k} \lambda} ( 0 ) 
 - \frac{ A_{\bd{k} \lambda} ( 0 ) }{ \omega_{\bd{k} \lambda}^2 }.
 \end{equation}

At this point we replace the spin operators
by classical vectors, ignoring the fact that different spin components do not commute.
The magnetoelastic field $\bd{F}_i ( t )$ can then be written as the
following functional of the spins,
 \begin{equation}
 \bd{F}_i ( t ) = \bar{\bd{h}}_i ( t ) + \delta \bd{h}_i ( t ) 
 - \int_0^t d t^\prime \sum_j 
 \mathbb{G}_{ij} ( t , t^\prime ) \dot{\bd{S}}_j ( t^\prime )  ,
 \end{equation}
where the part $\bar{\bd{h}}_i ( t )$ of the induced magnetic field
is independent of the phonon dynamics, 
 \begin{eqnarray}
  \bar{\bd{h}}_i ( t ) & = & \frac{1}{2 S^4 } \sum_{\alpha \beta} 
 \sum_{\mu \nu }  {\bd{e}}_{\alpha}   B_{\alpha \beta } B_{\mu \nu }
 \sum_j  S_i^{\beta} ( t ) S_j^{\mu} ( t ) S_j^{\nu} ( t ) 
 \nonumber
 \\
 &  \times & \frac{1}{N}  \sum_{\bd{k} \lambda} e^{ i \bd{k} \cdot ( \bd{R}_i - \bd{R}_j ) }
 \frac{ ( \bd{k}_{\alpha \beta} \cdot \bd{e}_{ \bd{k} \lambda} )
  ( \bd{k}_{\mu \nu} \cdot \bd{e}_{ -\bd{k} \lambda} )}{ M \omega_{\bd{k}\lambda}^2 },
 \nonumber
 \\
 & &
 \label{eq:barh}
 \end{eqnarray}
while the part $\delta {\bd{h}}_i (t)$
depends on the initial conditions of the phonons coordinates and momenta at time $t=0$,
 \begin{eqnarray}
\delta {\bd{h}}_i (t ) & = & - \frac{ i }{N S^2} \sum_{\alpha \beta} B_{\alpha \beta}
 {\bd{e}}_{\alpha} S_i^{\beta} (t ) \sum_{\bd{k} \lambda}
 e^{ i \bd{k} \cdot \bd{R}_i }
 \bd{k}_{\alpha \beta } \cdot \bd{e}_{\bd{k} \lambda} 
 \nonumber
 \\
 &  \times &
 \left[
 \tilde{X}_{\bd{k} \lambda} ( 0 ) \cos ( \omega_{\bd{k} \lambda} t )
 + \frac{ P_{\bd{k} \lambda} (0)}{M \omega_{\bd{k} \lambda} } 
 \sin ( \omega_{\bd{k} \lambda} t ) \right].
 \nonumber
 \\
 & &
 \label{eq:deltah}
 \end{eqnarray}
Finally, the damping kernel is given by
 \begin{eqnarray}
  \mathbb{G}^{\alpha \beta}_{ij} ( t , t^\prime )  & = & \frac{1}{S^4}
 \sum_{\mu \nu} B_{\alpha \mu} B_{ \beta \nu } S^{\mu}_i ( t ) S^\nu_j ( t^{\prime} )
 \nonumber
 \\
 & \times & \frac{1}{N}  \sum_{\bd{k} \lambda} e^{ i \bd{k} \cdot ( \bd{R}_i - \bd{R}_j ) }
  ( \bd{k}_{\alpha \mu} \cdot \bd{e}_{ \bd{k} \lambda} )
  ( \bd{k}_{ \beta \nu } \cdot \bd{e}_{ -\bd{k} \lambda} )
 \nonumber
 \\
 & \times &
 \frac{
 \cos [ \omega_{\bd{k} \lambda} ( t - t^{\prime} )]}{ M \omega_{\bd{k}\lambda}^2 }   .
 \hspace{7mm}
 \label{eq:GdefApp}
 \end{eqnarray}
Let us now assume that the (shifted) initial values $\tilde{X}_{\bd{k} \lambda} ( 0 )$ and
$P_{\bd{k} \lambda} ( 0 )$ of the normal modes
are classical Gaussian random variables with zero average and variance given by the
classical equipartition theorem with temperature $T$. 
The averages then vanish,
 $\langle \tilde{X}_{\bd{k} \lambda} (0) \rangle = 0
=
  \langle {P}_{\bd{k} \lambda} (0) \rangle $,
while the second moments are
 \begin{subequations}
 \begin{eqnarray}
  \langle \tilde{X}_{\bd{k} \lambda} (0)
\tilde{X}_{\bd{k}^{\prime} \lambda^{\prime}}  (0 ) \rangle& = &
 N \delta_{ \bd{k} , - \bd{k}^{\prime}} \delta_{\lambda \lambda^{\prime}}
  \frac{T} {M \omega_{\bd{k} \lambda}^2 }  ,
 \\
 \langle P_{\bd{k} \lambda} (0) P_{\bd{k}^{\prime} \lambda^{\prime}} (0) 
 \rangle & = & N \delta_{ \bd{k} , - \bd{k}^{\prime}} \delta_{\lambda \lambda^{\prime}}
  M T,
  \\
 \langle \tilde{X}_{\bd{k} \lambda} (0)
 {P}_{\bd{k}^{\prime} \lambda^{\prime}}  (0 ) \rangle& = & 0,
 \end{eqnarray}
\end{subequations}
where $\langle \ldots \rangle$ denotes the Gaussian averaging over the 
classical thermal phonon distribution.
Note that this implies
 \begin{equation}
 \langle \delta \bd{h}_i ( t ) \rangle =0,
 \end{equation}
and that the covariance of the random fields $\delta h^{\alpha}_i ( t )$ is related to the
matrix elements of the damping kernel
 $\mathbb{G}_{ij} ( t , t^\prime ) $
via the fluctuation-dissipation theorem,
 \begin{equation}
 \langle \delta h_i^{\alpha} ( t ) \delta h_j^{\beta} ( t^{\prime} )
 \rangle =  T  \mathbb{G}_{ij}^{\alpha \beta} ( t , t^\prime ) .
 \label{eq:FDT}
 \end{equation} 
With $ {\bd{h}}_i ( t )  = \bar{\bd{h}}_i ( t ) + \delta \bd{h}_i ( t )$,
we finally obtain the stochastic Landau-Lifshitz-Gilbert equation with non-Markovian
damping given in Eq.~($1$) of the main text, 
 \begin{eqnarray}
  \dot{\bd{S}}_i  ( t )  & = & \bd{S}_i ( t ) \times
 \Bigl[ {\bd{H}} ( t ) +  \bd{h}_i ( t ) + \sum_j \mathbb{K}_{ij} \bd{S}_j ( t ) 
 \Bigr]
 \nonumber
 \\
 &  - & 
 {\bd{S}}_i ( t ) \times  \int_0^t d t^\prime \sum_j 
 \mathbb{G}_{ij} ( t , t^\prime ) \dot{\bd{S}}_j ( t^\prime )   .
 \label{eq:LLGappendix}
 \end{eqnarray}
Assuming that the phonon dynamics is much faster than the
magnon dynamics, and that the damping kernel is local in the spatial indices and
diagonal
in the spin labels, we may approximate
 \begin{equation}
  \mathbb{G}_{ij}^{\alpha \beta}  ( t , t^\prime ) \approx 2 \gamma  
 \delta_{ij} \delta^{\alpha \beta}
 \delta ( t - t^{\prime} ).
 \label{eq:markov}
 \end{equation}
In this approximation Eq.~(\ref{eq:LLGappendix}) reduces to the usual form of the 
Markovian  LLG 
  \begin{eqnarray}
  \dot{\bd{S}}_i  ( t )  & = & \bd{S}_i ( t ) \times
 \Bigl[ {\bd{H}} ( t ) + {\bd{h}}_i ( t )
+ \sum_j \mathbb{K}_{ij} \bd{S}_j ( t ) 
 -  \gamma \dot{\bd{S}}_i (t)
 \Bigr].
 \nonumber
 \\
 & &
 \end{eqnarray}
However, because the characteristic phonon energies in YIG
have the same order of magnitude as the magnon energies,
the approximation (\ref{eq:markov}) is not sufficient
for a realistic description of the interplay between magnon and phonon dynamics
in YIG.

\section{Numerical solution of the LLG}
\label{sec:derivation}

\subsection{Reduction of the non-Markovian random process to Gaussian white noise}

In order to numerically solve the non-Markovian stochastic Landau-Lifshitz-Gilbert equation (\ref{eq:LLGappendix}) it is convenient
to reduce it to a Markovian equation containing only Gaussian white noise by introducing additional variables. 
The only approximation necessary is to neglect the dependence of the induced field and the damping kernel
on the state of the spin system by replacing $\bd{S}_i(t)\to S\bd{e}_z$ in their respective definitions, Eqs.~(\ref{eq:deltah}) and (\ref{eq:GdefApp}).
Then it is convenient to define the Fourier transforms
\begin{eqnarray}
 \delta\bd{h}_{\bd{k}}(t) & = & \sum_i e^{-i\bd{k}\cdot\bd{R}_{i}} \delta\bd{h}_i(t)
 = \sum_\lambda \delta\bd{h}_{\bd{k}\lambda} (t) , \\
 \mathbb{G}_{\bd{k}}(t,0) & = & \sum_i e^{-i\bd{k}\cdot\bd{R}_{ij}} \mathbb{G}_{ij}(t,0)
 = \sum_\lambda \tilde{\mathbb{G}}_{\bd{k}\lambda} \cos( \omega_{\bd{k}\lambda} t ) ,
 \nonumber
 \\
 & &
 \label{eq:gamma_k}
\end{eqnarray}
where 
\begin{eqnarray} 
 \delta {\bd{h}}_{\bd{k}\lambda} (t ) & = & - \frac{ i }{S} \sum_{\alpha} B_{\alpha z}
 {\bd{e}}_{\alpha} 
 \bd{k}_{\alpha z } \cdot \bd{e}_{\bd{k} \lambda} 
 \nonumber
 \\
 &  \times &
 \left[
 \tilde{X}_{\bd{k} \lambda} ( 0 ) \cos ( \omega_{\bd{k} \lambda} t )
 + \frac{ P_{\bd{k} \lambda} (0)}{M \omega_{\bd{k} \lambda} } 
 \sin ( \omega_{\bd{k} \lambda} t ) \right] , 
 \nonumber
 \\
 & & \\
 \tilde{\mathbb{G}}_{\bd{k}\lambda} &=& 
 B_{\alpha z}B_{\beta z}    
 \frac{ \left( \bd{k}_{\alpha z}\cdot\bd{e}_{\bd{k}\lambda} \right) \left( \bd{k}_{\beta z}\cdot\bd{e}_{-\bd{k}\lambda} \right) }
 { M \omega_{\bd{k}\lambda}^2 S^2 } .
 \label{eq:G}
\end{eqnarray}
The corresponding momentum space version of the fluctuation-dissipation theorem (\ref{eq:FDT}) is 
\begin{eqnarray}
 \langle \delta h_{\bd{k}\lambda}^\alpha(t)  
 \delta h_{\bd{k}^\prime\lambda'}^{\beta}(t^\prime) \rangle 
 &=&
 N \delta_{\bd{k},-\bd{k}^\prime} \delta_{\lambda\lambda'} T \tilde{\mathbb{G}}^{\alpha\beta}_{\bd{k}\lambda}
 \nonumber
 \\
 & & \times 
 \cos\left[ \omega_{\bd{k}\lambda}(t-t')\right] .
 \label{eq:FDT_klambda}
\end{eqnarray}
In the following we will neglect the average field $\bar{\bd{h}}_i(t)$ since it only gives rise to a small correction of the external field,
which we take as
\begin{equation}
 \bd{H}(t) = [ H_0 + H_1 \cos(2\omega_{\rm p}t) ] \bd{e}_z .
\end{equation}
Also, we will use that for thin YIG films the dipolar tensor is diagonal in the spin components \cite{Kreisel09s}.
Expanding the spins in the LLG (\ref{eq:LLGappendix}) as 
\begin{equation}
 \bd{S}_i(t) = S \left[ \bd{e}_z + \frac{1}{N} \sum_{\bd{k}} e^{i\bd{k}\cdot\bd{R}_i} \bd{m}_{\bd{k}}(t) \right]
\end{equation}
yields the equation of motion for the Fourier modes $\bd{m}_{\bd{k}}(t)$,
\begin{eqnarray}
 \dot{\bd{m}}_{\bd{k}} &=& 
 \left[ \mathbb{L}_{\bd{k}} + \mathbb{P}(t) \right] \bd{m}_{\bd{k}} 
 + \bd{e}_z \times \sum_\lambda \Bigl[  \delta\bd{h}_{\bd{k}\lambda}(t)  
 \nonumber\\
 &&- S\tilde{\mathbb{G}}_{\bd{k}\lambda} \int_0^t dt' 
 \cos\left[ \omega_{\bd{k}\lambda}(t-t')\right] \dot{\bd{m}}_{\bd{k}}(t')   \Bigr]
 \nonumber\\
 &&+ \frac{1}{N}\sum_{\bd{q}} \bd{m}_{\bd{k}-\bd{q}} \times \biggl\{
  S\mathbb{K}_{\bd{q}} \bd{m}_{\bd{q}} 
  + \sum_\lambda \Bigl[ \delta\bd{h}_{\bd{q}\lambda}(t) 
  \nonumber\\
 &&- S\tilde{\mathbb{G}}_{\bd{q}\lambda} \int_0^t dt' 
 \cos\left[ \omega_{\bd{q}\lambda}(t-t')\right] \dot{\bd{m}}_{\bd{q}}(t') \Bigr]
 \biggr\} .
 \nonumber
 \\
 & & 
  \label{eq:mDot}
\end{eqnarray}
Here we introduced the matrices
\begin{equation}
 \mathbb{L}_{\bd{k}} =
 \left( \begin{matrix}
       0 & \omega_{\bd{k}}^y & 0 \\
       - \omega_{\bd{k}}^x& 0 & 0 \\
       0 & 0 & 0
       \end{matrix} \right)
 , 
\end{equation}
and
\begin{equation}
 \mathbb{P}(t)= H_1 \cos(2\omega_{\rm p}t)
  \left( \begin{matrix}
       0 & 1 & 0 \\
       - 1 & 0 & 0 \\
       0 & 0 & 0
       \end{matrix} \right) .
\end{equation}
$\mathbb{L}_{\bd{k}}$ describes the linear oscillations of the spins
in terms of the frequencies 
 \begin{eqnarray}
  \omega_{\bd{k}}^x &=& H_0 + S \left( \mathbb{K}_{\bd{k}=0}^{zz} - \mathbb{K}_{\bd{k}}^{xx}  \right) , \\
  \omega_{\bd{k}}^y &=&  H_0 + S \left( \mathbb{K}_{\bd{k}=0}^{zz} - \mathbb{K}_{\bd{k}}^{yy}  \right) ,
 \end{eqnarray}
whereas $\mathbb{P}(t)$ contains the parametric pumping.
Note that the nonvanishing eigenvalue of the matrix $\mathbb{L}_{\bd{k}}$
gives the magnon dispersion of linear spin wave theory,
\begin{equation}
 E_{\bd{k}} = \sqrt{ \omega_{\bd{k}}^x \omega_{\bd{k}}^y } .
\end{equation}

The simple harmonic time dependence of the damping tensor (\ref{eq:gamma_k}) now allows us to convert this non-Markovian 
equation of motion into a system of Markovian ones \cite{Farias09as,Farias09bs}. To this end we introduce the new set of variables
 \begin{align}
  \bd{u}_{\bd{k}\lambda} =& S\tilde{\mathbb{G}}_{\bd{k}\lambda}  
  \int_0^t dt'\cos\left[ \omega_{\bd{k}\lambda}(t-t')\right] \dot{\bd{m}}_{\bd{k}}(t') 
  \nonumber\\
  &- S\tilde{\mathbb{G}}_{\bd{k}\lambda} \bd{m}_{\bd{k}}  .
  \label{eq:Def_U}
 \end{align}
 They satisfy the second order equations of motion
 \begin{equation}
   \ddot{\bd{u}}_{\bd{k}\lambda} = -\omega_{\bd{k}\lambda}^2 \left[ \bd{u}_{\bd{k}\lambda} + S\tilde{\mathbb{G}}_{\bd{k}\lambda} \bd{m}_{\bd{k}} \right] ,
  \label{eq:Uddot}
 \end{equation}
 with initial conditions
 \begin{equation}
  \bd{u}_{\bd{k}\lambda}(0) = - S\tilde{\mathbb{G}}_{\bd{k}\lambda} \bd{m}_{\bd{k}}(0),
  \;\;\;
  \dot{\bd{u}}_{\bd{k}\lambda}(0)=0.
 \end{equation}
 With the definition (\ref{eq:Def_U}) of the new variables the spin equation of motion (\ref{eq:mDot}) becomes
 \begin{eqnarray}
 \dot{\bd{m}}_{\bd{k}} &=& 
 \left[ \mathbb{L}_{\bd{k}} + \mathbb{P}(t) \right] \bd{m}_{\bd{k}} 
 \nonumber\\
 &&
 + \bd{e}_z \times \sum_\lambda \Bigl[  \delta\bd{h}_{\bd{k}\lambda}(t)  
 - \bd{u}_{\bd{k}\lambda} - S\tilde{\mathbb{G}}_{\bd{k}\lambda} \bd{m}_{\bd{k}}   \Bigr]
 \nonumber\\
 &&+ \frac{1}{N}\sum_{\bd{q}} \bd{m}_{\bd{k}-\bd{q}} \times \biggl\{
  S\mathbb{K}_{\bd{q}} \bd{m}_{\bd{q}} 
 \nonumber\\
 &&
  + \sum_\lambda \Bigl[ \delta\bd{h}_{\bd{q}\lambda}(t) 
  - \bd{u}_{\bd{q}\lambda} - S\tilde{\mathbb{G}}_{\bd{q}\lambda} \bd{m}_{\bd{q}} \Bigr]
 \biggr\} .
  \label{eq:USdot}
\end{eqnarray}

The remaining task is to simulate a Gaussian random process with covariance given by the 
fluctuation-dissipation theorem (\ref{eq:FDT_klambda}). As will be shown below, such a process can be modeled with 
a driven oscillator equation of motion \cite{Munakata85s,Farias09as,Farias09bs}
\begin{equation}
 \left( \partial_t^2 + 2g\omega_{\bd{k}\lambda} \partial_t 
 + \omega_{\bd{k}\lambda}^2 \right) \delta\bd{h}_{\bd{k}\lambda} =
 2 \sqrt{gT\omega_{\bd{k}\lambda}^3} \bd{b}_{\bd{k}\lambda} f_{\bd{k}\lambda}(t),
 \label{eq:Hddot}
\end{equation}
where the driving term $f_{\bd{k}\lambda}(t)$ is a complex Gaussian white noise process
with vanishing average and covariance
\begin{equation}
 \langle f_{\bd{k}\lambda}(t) f_{\bd{k}'\lambda'}^*(t') \rangle 
 = N\delta_{\bd{k}\bd{k}'} \delta_{\lambda\lambda'} \delta(t-t') .
\end{equation}
The coupling vector is given by
\begin{equation}
 \bd{b}_{\bd{k}\lambda} = \sum_\alpha \bd{e}_\alpha B_{\alpha z}
 \frac{ i \bd{k}_{\alpha z}\cdot \bd{e}_{\bd{k}\lambda} }{ \sqrt{M}\omega_{\bd{k}\lambda} S },
\end{equation}
so that $b_{\bd{k}\lambda}^\alpha b_{-\bd{k}\lambda}^{\beta} = \tilde{\mathbb{G}}_{\bd{k}\lambda}^{\alpha\beta}$.
We also added a damping term with strength $g < 1$ which is necessary to obtain a finite covariance function
and can be interpreted as the damping of the phonon subsystem.
In the end we have to take the limit $g\to 0$ to reproduce the Fluctuation-Dissipation theorem (\ref{eq:FDT_klambda})
from the driven oscillator (\ref{eq:Hddot}). 
To stay consistent, a corresponding term $-2g\omega_{\bd{k}\lambda} \dot{\bd{u}}_{\bd{k}\lambda}$ should also be added on the right-hand side
of the equation of motion (\ref{eq:Uddot}) for $\bd{u}_{\bd{k}\lambda}$.

We have now succeded in our task to get rid of the non-Markovian terms and the colored noise, at the expense
of having introduced two new sets of equations of motion. However, it turns out that we can combine these
two sets into a single one since they are both linear oscillator equations and the spin equation of motion (\ref{eq:USdot})
depends only on the difference
\begin{equation}
 \bd{v}_{\bd{k}\lambda} = \delta\bd{h}_{\bd{k}\lambda} - \bd{u}_{\bd{k}\lambda} ,
\end{equation}
which also obeys a driven oscillator equation. Therefore we are left with the two coupled sets of 
stochastic differential equations
\begin{subequations} \label{eq:sLLGfinal}
\begin{eqnarray}  
 \dot{\bd{m}}_{\bd{k}} &=& 
 \left[ \mathbb{L}_{\bd{k}} + \mathbb{P}(t) \right] \bd{m}_{\bd{k}} 
 \nonumber\\
 &&+ \bd{e}_z \times \sum_\lambda \left[  \bd{v}_{\bd{k}\lambda}   - 
 S\tilde{\mathbb{G}}_{\bd{k}\lambda} \bd{m}_{\bd{k}}  \right]
 \nonumber\\
 &&+ \frac{1}{N}\sum_{\bd{q}} \bd{m}_{\bd{k}-\bd{q}} \times \biggl\{
  S\mathbb{K}_{\bd{q}} \bd{m}_{\bd{q}} \nonumber\\
  && \phantom{+\frac{1}{N}\sum_{\bd{q}}}
  + \sum_\lambda \left[ \bd{v}_{\bd{q}\lambda}   - 
 S\tilde{\mathbb{G}}_{\bd{q}\lambda} \bd{m}_{\bd{q}} \right]
 \biggr\} , 
 \label{eq:sLLGa}\\
 \ddot{\bd{v}}_{\bd{k}\lambda} &=& -2g\omega_{\bd{k}\lambda} \dot{\bd{v}}_{\bd{k}\lambda}
 - \omega_{\bd{k}\lambda}^2
 \left[ \bd{v}_{\bd{k}\lambda} - S\tilde{\mathbb{G}}_{\bd{k}\lambda} \bd{m}_{\bd{k}} \right]
 \nonumber\\
 &&+ 2 \sqrt{gT\omega_{\bd{k}\lambda}^3} \bd{b}_{\bd{k}\lambda} f_{\bd{k}\lambda}(t) .
\end{eqnarray}
\end{subequations}
with complex white noise $f_{\bd{k}\lambda}(t)$, and the constraint
\begin{equation}
  \bd{v}_{\bd{k}\lambda}(0) = S\tilde{\mathbb{G}}_{\bd{k}\lambda} \bd{m}_{\bd{k}}(0) ,
  \;\;\;
  \dot{\bd{v}}_{\bd{k}\lambda}(0)=0 
\end{equation}
on the initial conditions of the auxiliary variables.

\subsection{Numerical implementation}

\begin{figure}[t]
\includegraphics[width=80mm]{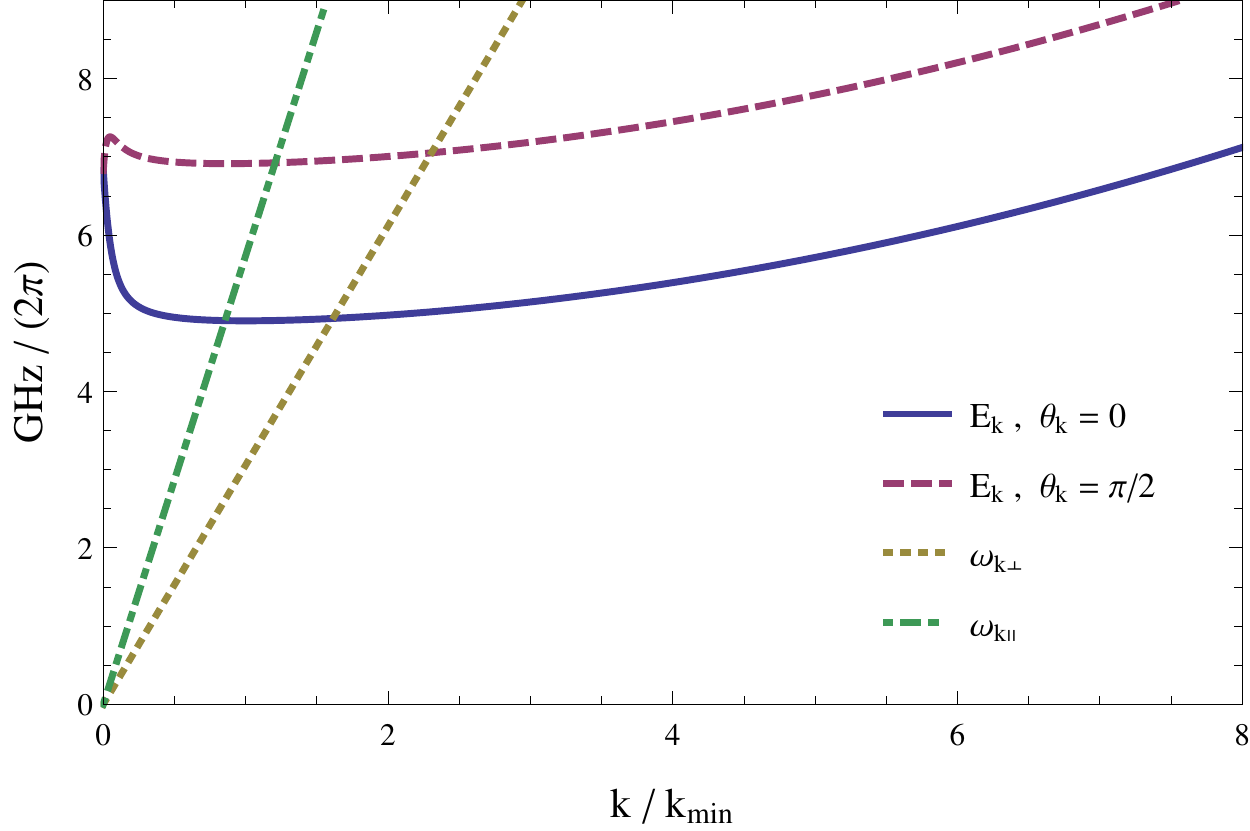}
  \caption{%
(Color online).
Dispersions of the magnon and phonon modes of a thin YIG stripe with
thickness $d = 6.7 \, {\rm \mu m}$ in an external magnetic field $H_0 =1710 \, {\rm Oe} \times \mu $, for $\bd{k}$ parallel and perpendicular to the
in-plane magnetic field. 
}
  \label{fig:modes}
\end{figure}

\begin{figure}[t]
\includegraphics[width=80mm,height=80mm]{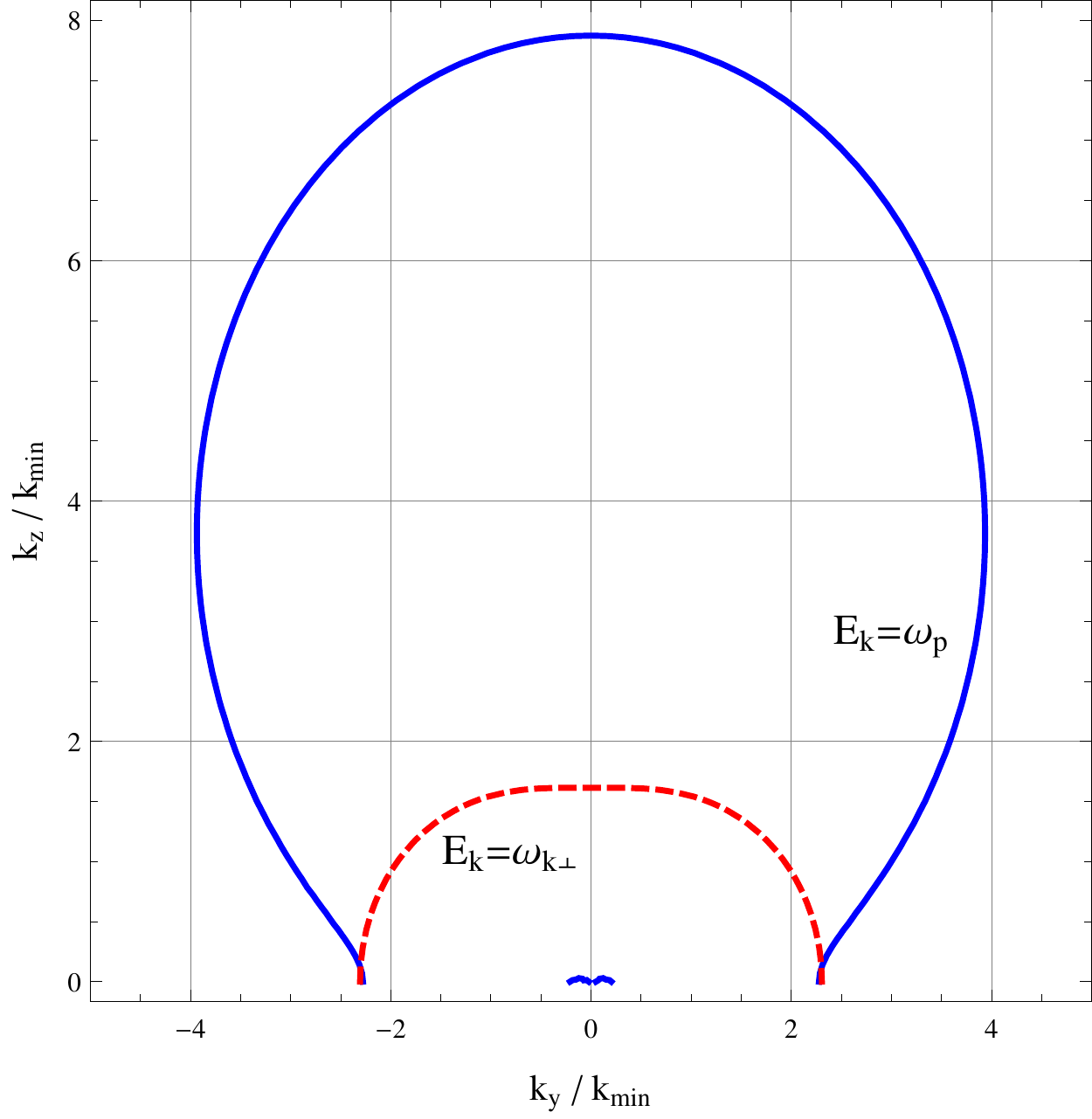}
  \caption{%
(Color online).
Momentum space contours of the energy surfaces of parametric resonance $E_{\bd{k}}=\omega_{\rm p}$ (blue solid line) 
and transverse magnetoelastic hybridization $E_{\bd{k}}=\omega_{\bd{k}\bot}$ (red dashed line).
}
  \label{fig:contour}
\end{figure}

A plot of the magnon and phonon dispersions for YIG with parameters mentioned in the main text is given in Fig. \ref{fig:modes}. 
Especially note that both phonon branches cross the magnon dispersion relatively close to the minimum of the magnon dispersion.
Therefore phonon and magnon dynamics occur at the same time scale and we cannot approximate the damping tensor (\ref{eq:GdefApp}) by 
white noise as in the conventional Markovian LLG.
However we will neglect the longitudinal phonon branch entirely and only consider the two transverse branches.
This can be justified with the fact that 
the damping tensor (\ref{eq:GdefApp}) is proportional to the square of the inverse sound velocity.
Therefore the faster longitudinal phonons are considerably less important for the magnon relaxation than the transverse phonons.
A plot of the corresponding energy surface $E_{\bd{k}}=\omega_{\bd{k}\bot}$ is shown in Fig. \ref{fig:contour},
together with the contour satisfying the parametric resonance condition $E_{\bd{k}}=\omega_{\rm p}$. 
Note however that the parametric resonance is only effective for elliptic spin waves which have momenta close to $k_z=0$ in thin YIG films.
The phonon damping coefficient necessary for the noise simulation will be set to $g=10^{-3}$.

For the explicit numerical solution we use the stochastic Heun scheme \cite{Palacios98s,Alvarez11s} to solve the system (\ref{eq:sLLGfinal}) of stochastic differential
equations on an equidistant momentum grid with increment $\Delta k = 0.2\,k_{\rm min}$. 
As we are mainly interested in the long-wavelength regime we
work with a truncated momentum space with $125\times 189=23625$ points, such that $|k_y|\le 12.4\,k_{\rm min}$ and $|k_z|\le 18.8\,k_{\rm min} $.
The integrations are discretized as 
\begin{equation}
 \frac{1}{N}\sum_{\bd{k}} = \left(\frac{a}{2\pi}\right)^2 \int d^2 k \approx \left(\frac{a}{2\pi}\right)^2 \sum_{\bd{k}} (\Delta k)^2
 = \frac{1}{N_{\rm eff}} \sum_{\bd{k}} ,
\end{equation}
where we defined the effective lattice size $N_{\rm eff}= 4\pi^2/(a\Delta k)^2 \approx 2.6\times 10^7 $.
The corresponding delta distributions are replaced according to $N\delta_{\bd{k}\bd{k}'}\to N_{\rm eff}\delta_{\bd{k}\bd{k}'}$ since $\frac{1}{N}\sum_{\bd{k}} N\delta_{\bd{k}\bd{k}'}=1$ has to hold.
As all momentum integrations are convolutions, they are performed using a fast Fourier transform algorithm \cite{NAGs}. 
Gaussian random numbers are created via the polar Box-Muller algorithm \cite{Thomas07s} from random numbers distributed
uniformly between $0$ and $1$. 
We calculate the stochastic time evolution of $\bd{m}_{\bd{k}}(t)$ from $t=0$ up to $t=800\,{\rm ns}$ in increments of $\Delta t = 10^{-4}\, {\rm ns}$, 
starting from a fully polarized ferromagnetic state with $\bd{m}_{\bd{k}}(0)=0$ for all momenta $\bd{k}$.

\section{Noise driven harmonic oscillator}
\label{App:ho}

This appendix will be devoted to simulating complex colored noise satisfying
\begin{eqnarray}
 \langle x(t) \rangle & = & 0 , \\
 \langle x(t+\tau) x^*(t) \rangle & = & K \cos(\Omega \tau) ,
\end{eqnarray}
via an underlying white noise process.
To this end we consider an (under-)damped, complex harmonic oscillator driven by Gaussian white noise, i.e.,
\begin{equation} \label{dampedHO}
 \left(\partial_t^2 + 2\gamma\partial_t + \Omega^2 \right) x = A f(t),
\end{equation}
with
\begin{eqnarray}
 \langle f(t) \rangle & = & 0 , \\
 \langle f(t+\tau) f^*(t) \rangle & = & \delta(\tau) .  
\end{eqnarray}
The general solution of Eq. (\ref{dampedHO}) in the underdamped situation is given by
\begin{equation}
 x(t) = x_h(t) + A \int dt' G(t-t') f(t'), 
\end{equation}
with the Green function 
\begin{equation}
 G(t) = \Theta(t) e^{-\gamma t} \frac{\sin\left(\sqrt{\Omega^2-\gamma^2} t\right)}{\sqrt{\Omega^2-\gamma^2}} ,
\end{equation}
and where $x_h(t)$ denotes the solution of the homogeneous equation. 
As we aim to simulate a stochastic process with vanishing average, we set $x_h(t)=0$ by choosing $x(0)=0=\dot{x}(0)$.
In frequency domain, we then obtain
\begin{equation}
 x(\omega) =  \int dt e^{i\omega t} x(t) = A  G(\omega) f(\omega) 
\end{equation}
where
\begin{equation}
 G(\omega) = \frac{1}{\Omega^2-\omega^2-2i\gamma\omega} .
\end{equation}
For the random force, we have
\begin{eqnarray}
 \langle f(t+\tau) f^*(t) \rangle 
 &=& \delta(\tau) \nonumber \\
 &=& \int \frac{\text{d}\omega_1}{2\pi} \int \frac{\text{d}\omega_2}{2\pi} e^{-i\omega_1(t+\tau)+i\omega_2t}
 \nonumber\\
 && \times \langle f(\omega_1) f^*(\omega_2) \rangle ,
\end{eqnarray}
and therefore
\begin{equation}
 \langle f(\omega_1) f^*(\omega_2) \rangle = 2\pi \delta(\omega_1-\omega_2).
\end{equation}
With the above, we can calculate the correlation function
\begin{subequations}
\begin{eqnarray}
 \langle x(t+\tau) x^*(t) \rangle 
 %
 %
 %
 &=& |A|^2 \int \frac{\text{d}\omega}{2\pi} e^{-i\omega\tau} | G(\omega) |^2 \\
 &=& |A|^2 \int \frac{\text{d}\omega}{2\pi} \frac{ e^{-i\omega\tau} }{ \left( \Omega^2-\omega^2 \right)^2 + 4\gamma^2\omega^2 } \nonumber\\&&\\
 &=& \frac{|A|^2}{4\gamma\Omega^2} e^{-\gamma|\tau|} \Biggl[ \cos\left(\sqrt{\Omega^2-\gamma^2} \tau\right) 
  \nonumber\\
 &&
 + \frac{\gamma}{\sqrt{\Omega^2-\gamma^2}} \sin\left(\sqrt{\Omega^2-\gamma^2} |\tau|\right)  \Biggr] .
 \nonumber\\&&
\end{eqnarray}
\end{subequations}
For our application, we are mainly interested in the limit of vanishing damping. To this end we define the dimensionless damping constant $g=\gamma/\Omega$, so that
\begin{eqnarray}
 \langle x(t+\tau) x^*(t) \rangle
 &=& \frac{|A|^2}{4g\Omega^3} e^{-g\Omega|\tau|} \Biggl[ \cos\left(\sqrt{1-g^2} \Omega \tau\right) 
 \nonumber\\
 &&+ \frac{g \Omega}{\sqrt{1-g^2}|\Omega|} \sin\left(\sqrt{1-g^2} |\Omega \tau|\right)  \Biggr] .\nonumber\\&&
\end{eqnarray}
Next, we demand that
\begin{equation}
 \lim_{g\to 0} \langle x(t+\tau) x^*(t) \rangle = K \cos(\Omega \tau) ,
\end{equation}
yielding
\begin{equation}
 |A| = \sqrt{4g\Omega^3 K} .
\end{equation}

\section{Parametric instability}

In this appendix we will derive the parametric instability from linear spin wave theory. Therefore we consider only the first line of 
the equation of motion (\ref{eq:sLLGa}) for the spin deviation, which explicitly yields
\begin{subequations} \label{PI:psiDotElliptic}
 \begin{align}
  \dot{m}^x_{\bd{k}} =& \left[ \omega_{\bd{k}}^y + H_1\cos(2\omega_{\rm p}t) \right] m^y_{\bd{k}} , \\
  \dot{m}^y_{\bd{k}} =& -\left[ \omega_{\bd{k}}^x + H_1\cos(2\omega_{\rm p}t) \right] m^x_{\bd{k}} . 
 \end{align}
\end{subequations}
The parametric instability is usually derived for the complex normal mode of the system without pumping, given by
\begin{equation} \label{PI:NormalModes}
 \phi_{\bd{k}} = \left(\frac{\omega_{\bd{k}}^x}{\omega_{\bd{k}}^y}\right)^{1/4} m^x_{\bd{k}} 
 + i\left(\frac{\omega_{\bd{k}}^y}{\omega_{\bd{k}}^x}\right)^{1/4} m^y_{\bd{k}} .
\end{equation} 
In terms of this normal mode the two linearized equation of motions (\ref{PI:psiDotElliptic}) become the single complex equation
\begin{eqnarray}
 \dot{\phi}_{\bd{k}} =& -i \left[ E_{\bd{k}} + U_{\bd{k}} H_1 \cos(2\omega_{\rm p}t) \right] \phi_{\bd{k}} 
 \nonumber\\
 & - 2 i V_{\bd{k}} H_1 \cos(2\omega_{\rm p}t) \phi^*_{-\bd{k}} ,
 \label{PI:phiDot}
\end{eqnarray}
where $U_{\bd{k}}=\sqrt{1+4|V_{\bd{k}}|^2}$ and
$V_{\bd{k}}=(\omega_{\bd{k}}^y-\omega_{\bd{k}}^x)/(4 E_{\bd{k}})=S (D_{\bd{k}}^{xx}-D_{\bd{k}}^{yy})/(4 E_{\bd{k}})$ is proportional to the ellipticity of the spin wave.
Next let us move to a reference frame rotating with the pumping frequency by setting
\begin{equation}
 \phi_{\bd{k}}(t) = e^{-i\omega_{\rm p}t} \varphi_{\bd{k}}(t).
\end{equation}
Inserting this ansatz into the equation of motion (\ref{PI:phiDot}) yields
\begin{eqnarray}
 \dot{\varphi}_{\bd{k}} &=& -i \left[ ( E_{\bd{k}} -\omega_{\rm p} ) + U_{\bd{k}} H_1 \cos(2\omega_{\rm p}t) \right] \varphi_{\bd{k}} 
 \nonumber\\
 && - i V_{\bd{k}} H_1 \left( e^{4i\omega_{\rm p}t} + 1 \right) \varphi^*_{-\bd{k}} .
 \label{PI:phiDotInt}
\end{eqnarray}
To study the parametric instability it is now sufficient to only retain the resonant term and drop all oscillating contributions in the above equation \cite{Gurevich96s}.
This leaves us with
\begin{equation}
  \dot{\varphi}_{\bd{k}} =  -i  ( E_{\bd{k}} -\omega_{\rm p} ) \varphi_{\bd{k}}
  - i V_{\bd{k}} H_1 \varphi^*_{-\bd{k}} .
 \label{PI:phiDotIntRot}
\end{equation}
Taking an additional time derivative turns this into
\begin{equation}
  \ddot{\varphi}_{\bd{k}} = - \left( | E_{\bd{k}} -\omega_{\rm p} |^2 - |V_{\bd{k}} H_1 |^2 \right)  \varphi_{\bd{k}} ,
\end{equation}
showing that the system becomes unstable for
\begin{equation}
 | E_{\bd{k}} -\omega_{\rm p} | < | V_{\bd{k}} H_1 | .
\end{equation}
This is the parametric instability \cite{Gurevich96s}. Note that it can only occur for elliptic spin waves, which are characterized by nonvanishing $V_{\bd{k}}$. 
It is also readily seen that the unstable modes diverge as $\varphi_{\bd{k}} \propto e^{\alpha t}$ ,
with
\begin{equation}
 \alpha_{\bd{k}}^2 = | V_{\bd{k}} H_1 |^2 - | E_{\bd{k}} -\omega_{\rm p} |^2 .
\end{equation}

\section{Thermal equilibrium}
\label{App:Eq}

To describe thermal equilibrium in a ferromagnetic state we may safely assume that the $x$ and $y$ components of the spin deviation vector $\bd{m}_i=(\bd{S}_i-S\bd{e}_z)/S$ 
are small compared to unity. Then we may expand
\begin{equation}
 m_i^z = \sqrt{1 - (m_i^x)^2 - (m_i^y)^2 } - 1 \approx - \frac{1}{2}\left[ (m_i^x)^2 + (m_i^y)^2 \right] .
\end{equation}
Retaining only terms quadratic in $m_i^x$ and $m_i^y$ in the Hamiltonian (\ref{eq:Hspin2}) and dropping the pumping field leads to
\begin{eqnarray}
 {\cal H}_{\rm m} & \approx & -SNH_0 - \frac{S^2}{2} \sum_{ij} \mathbb{K}_{ij}^{zz} + \frac{S}{2} \sum_{ij} \sum_{\alpha=x,y} \nonumber\\
 && \times \left[ \delta_{ij} \left( H_0 + S\sum_n \mathbb{K}_{in}^{zz} \right) - S\mathbb{K}_{ij}^{\alpha\alpha} \right] m_i^\alpha m_j^\alpha \nonumber\\
 &=& -SNH_0 - \frac{S^2}{2} \mathbb{K}_{\bd{k}=0}^{zz} + \frac{S}{2N} \sum_{\bd{k}} \sum_{\alpha=x,y} \omega_{\bd{k}}^{\alpha} m_{-\bd{k}}^\alpha m_{\bd{k}}^\alpha . 
 \nonumber\\&&
\end{eqnarray}
As we are now restricting ourselves to a noninteracting system, it is again convenient to introduce the normal modes (\ref{PI:NormalModes}). In terms of them
the above Hamiltonian is given by
\begin{equation}
 {\cal H}_{\rm m} = -SNH_0 - \frac{S^2}{2} \mathbb{K}_{\bd{k}=0}^{zz} + \frac{S}{2N} \sum_{\bd{k}} E_{\bd{k}} \phi_{\bd{k}}^* \phi_{\bd{k}} ,
\end{equation}
which is just a sum of ground state energy and an ensemble of noninteracting, classical harmonic oscillators. In thermal equilibrium each of them has to satisfy the
equipartition theorem
\begin{equation}
 \langle \phi_{\bd{k}}^* \phi_{\bd{k}^\prime} \rangle = N \delta_{\bd{k}\bd{k}^\prime} \frac{2 T}{S E_{\bd{k}} } ,
\end{equation}
and therefore the thermal equilibrium value of the square of the transverse spin component $n_{\bd{k}}=|m_{\bd{k}}^x|^2+|m_{\bd{k}}^y|^2$ is given by
\begin{equation}
 n_{\bd{k}}^{\rm th} =  \left\langle  n_{\bd{k}} \right\rangle = N \sqrt{1+4|V_{\bd{k}}|^2} \frac{2 T}{SE_{\bd{k}}}  .
\end{equation}

\end{document}